\documentclass[pdflatex,sn-mathphys-num]{sn-jnl}
\usepackage{graphicx}
\usepackage{multirow}
\usepackage{amsmath,amssymb,amsfonts}
\usepackage{amsthm}
\usepackage{mathrsfs}
\usepackage[title]{appendix}
\usepackage{xcolor}
\usepackage{textcomp}
\usepackage{manyfoot}
\usepackage{booktabs}
\usepackage{algorithm}
\usepackage{algorithmicx}
\usepackage{algpseudocode}
\usepackage{listings}
\usepackage{subfigure}
\usepackage{clipboard}
\theoremstyle{thmstyleone}
\newtheorem{theorem}{Theorem}

\theoremstyle{thmstyletwo}

\newtheorem{remark}{Remark}
\theoremstyle{thmstylethree}

\newtheorem{corollary}{Corollary}
\newtheorem{lemma}{Lemma}
\newcommand{\qsj}[1]{\textcolor[rgb]{0.00,0.00,0.00}{#1}}

\begin{document}

\title[Article Title]{Comparative Statics of Trading Boundary in Finite-horizon Portfolio Selection Problem with Proportional Transaction Costs}

\author*[1]{\fnm{Shuaijie} \sur{Qian}}\email{sjqian@ust.hk}

\author[2]{\fnm{Jintao} \sur{Li}}\email{e0622340@u.nus.edu}
\equalcont{These authors contributed equally to this work.}

\affil[1]{\orgdiv{Department of Mathematics}, \orgname{The Hong Kong University of Science and Technology}, \orgaddress{\street{Clear Water Bay}, \city{ Hong Kong}}}

\affil[2]{\orgdiv{Finance and Financial Risk Management}, \orgname{NUS (Chongqing) Research Institute}, \orgaddress{\street{ Chongqing Liang Jiang New Area}, \city{Chongqing}, \postcode{401123}, \country{ China}}}

\abstract{We consider Merton's problem with proportional transaction costs. It is well known that the optimal investment strategy is characterized by two trading boundaries—the buy boundary and the sell boundary—between which lies the no-trading region. We investigate how these two trading boundaries vary with the transaction cost rates. We show that the cost-adjusted trading boundaries are monotone in the transaction costs. Our result implies the following: (i) the Merton line must lie between the two cost-adjusted trading boundaries; and (ii) when the Merton line is positive, both the buy and sell boundaries are monotone in the transaction cost rates, and consequently the Merton line lies in the no-trading region.}

\keywords{HJB equation, Transaction costs, Free boundary, Singular control}

\pacs[JEL Classification]{C61}

\maketitle

\section{Introduction}\label{sec1}

\citet{MERTON1971373} pioneers the study of continuous-time portfolio selection and shows that in the absence of transaction costs, a CRRA investor with access to one stock and one bond should optimally maintain a constant bond--stock ratio, referred to as the Merton line. \citet{MAGILL1976245} incorporate transaction costs into Merton's model and find that a no-trading region emerges: trading occurs only when the bond--stock ratio hits the boundary of this region, i.e., the trading boundary. The trading boundary consists of a buy boundary and a sell boundary. When the ratio exceeds the buy boundary (falls below the sell boundary), the investor should purchase (sell) stock to keep the ratio within the no-trading region.

In this paper, {our objective is to explore how the trading boundary varies with transaction costs in the finite-horizon problem.} Intuitively, one may conjecture that the buy (sell) boundary should be increasing (decreasing) in the transaction cost rates. In other words, the no-trading region under lower costs would be contained within the one under higher costs. While this conjecture is consistent with the fact that the Merton line lies within the no-trading region in the no-leverage case, it conflicts with the finding of \citet{Shreve1994} that the Merton line is likely to lie beyond the no-trading region when leverage is {optimal}.

Using a rigorous mathematical analysis, we show that in the no-leverage case, the trading boundaries are indeed monotone with respect to transaction costs, meaning that the no-trading region expands (shrinks) when transaction costs increase (decrease). \qsj{However, the conclusion may fail to hold when leverage is optimal, as evidenced by the finding of \citet{Shreve1994} aforementioned.}

The main contribution of this paper is to {establish the following general result}: the trading boundary {adjusted} by transaction costs is monotone in the transaction cost rates. More {precisely}, we prove that the sell boundary adjusted by the sale cost rate and the buy boundary adjusted by the purchase cost rate, i.e.,
\[
\frac{x_s(t)}{1-\mu} \quad \text{and} \quad \frac{x_b(t)}{1+\lambda},
\]
are monotone decreasing and increasing, respectively, as the cost rates $\mu$ and $\lambda$ increase. Here $x_s(t)$ and $x_b(t)$ denote the sell and buy boundaries, while $\mu\in[0,1)$ and $\lambda\in[0,\infty)$ are the proportional transaction cost rates for selling and buying the stock, respectively. This monotonicity leads to several implications:
\begin{enumerate}
    \item[(i)] For a fixed sale (purchase) cost rate, the sell (buy) boundary is monotone in the purchase (sale) cost rate.
    \item[(ii)] When leverage is {not} optimal (i.e., $x_b(t), x_s(t)\ge 0$), the boundaries $x_b(t)$ and $x_s(t)$ are monotone with respect to transaction cost rates.
    \item[(iii)] When the optimal strategy {involves} leverage (i.e., $x_b(t)\le 0$ or $x_s(t)\le 0$), the buy (sell) boundary need not be monotone in the purchase (sale) cost rate.
\end{enumerate}

This paper focuses on the finite-horizon problem with logarithmic utility and consumption. {When replacing the $\log$ utility with a general power utility, our technique works \qsj{for the case} without consumption, although we believe the results remain valid \qsj{if consumption is included}.}

Technically, we reformulate the problem by merging the purchase and sale cost rates into a single parameter and then apply a comparison principle for the associated PDE of the equivalent double obstacle problem. This approach also applies to studying how the trading boundary depends on other model parameters—such as the risk premium, risk-aversion level, and stock volatility—for the finite-horizon problem without consumption.

\citet{Davis1990} and \citet{Shreve1994} provide rigorous theoretical analyses of the trading boundaries. \citet{Loewenstein2002}, \citet{DAI20091445}, and \citet{Dai2009Consumption} investigate their properties in finite-horizon settings. {\citet{Hobson2019} focus on the infinite-horizon problem and prove a similar dependency of the trading boundary on transaction costs via a semi-closed-form representation of the value function obtained from the corresponding ODE. However, this approach works only for the one-dimensional infinite-horizon case and therefore cannot handle our finite-horizon setting.} \qsj{The asymptotic analysis in \citet{janevcek2004asymptotic} indicates that in the infinite-horizon problem, the Merton line lies within the no-trading region when transaction cost rates are small. However, for the finite-horizon counterpart, if the Merton line is negative, then regardless of the cost rates, it lies outside the no-trading region when the time to maturity is sufficiently short.}

{The remainder of the paper is organized as follows. Section \ref{sec model} introduces the model setup. Section \ref{Sec mono cost} develops the theoretical analysis. \qsj{Section \ref{sec numerical} provides the numerical analysis.} Section \ref{sec conclusion} concludes.} \qsj{Additional theoretical and numerical results are relegated to the Appendix.}

\section{Model setup}\label{sec model}

\subsection{The market}
We assume that the investor has access to two assets: a risky stock and a risk-free bond with interest rate $r > 0$. The stock price follows the stochastic differential equation
\begin{equation*}
	\begin{aligned}
		d S_{ t} &= S_{t}\left(\alpha\, dt + \sigma\, d \mathcal{B}_{t}\right),
	\end{aligned}
\end{equation*}
where $\alpha$ is the expected return, $\sigma$ is the volatility, and $\{\mathcal{B}_{t}\}_{t\geq 0}$ is a standard one-dimensional Brownian motion on a filtered probability space $\big(\mathbb{S}, \mathscr{F}, \{\mathscr{F}_{t}\}_{t\geq 0}, P\big)$ with $\mathcal{B}_{0}=0$ almost surely. The filtration $\{\mathscr{F}_{t}\}_{t\geq 0}$ is generated by $\{\mathcal{B}_{t}\}_{t\geq 0}$, which is right-continuous, and $\mathscr{F}_{t}$ contains all $P$-null sets of $\mathscr{F}$ for any $t\geq 0$. Throughout the paper, we assume $\alpha > r$.

Trading the stock incurs proportional transaction costs. Let $X_t$ and $Y_t$ denote the dollar amounts invested in the bond and stock accounts, respectively. Their dynamics are given by
\begin{align}
	dX_t &= \big(rX_t - c_t\big) \,dt - (1+\lambda) \,dL_t + (1-\mu) \,dM_t, \label{equ evolution1}\\
	dY_t &= \alpha Y_t \,dt + \sigma Y_t \,d\mathcal{B}_t + dL_t - dM_t, \label{equ evolution2}
\end{align}
where $c_t \ge 0$ is the consumption rate, and $L_t$ and $M_t$ denote the cumulative dollar amounts of stock purchases and sales, respectively. Both $L_t$ and $M_t$ are right-continuous with left limits, non-negative, non-decreasing, and adapted to $\{\mathscr{F}_t\}_{t\geq 0}$. The parameters $\lambda \in [0,+\infty)$ and $\mu \in [0, 1)$ represent the proportional transaction cost rates on purchases and sales, respectively. To avoid the problem from degenerating into the classical Merton's problem, we require $\lambda + \mu > 0$.

\subsection{The investor's problem}

Since $\alpha > r$, short-selling the stock is never optimal. Hence, we set $Y_t \ge 0$ for all $0 \le t \le T$, \qsj{where $T$ is the terminal time}. Let $W_t$ denote the investor's net wealth at time $t$, defined by
\[
	W_t = X_t + (1-\mu) Y_t .
\]
To ensure that net wealth remains non-negative, the solvency region is defined as
\[
	\mathscr{S} = \left\{ (x, y) \in \mathbb{R}^2 : x + (1-\mu) y \ge 0,\; y \ge 0 \right\}.
\]

Given an initial position $(x, y) \in \mathscr{S}$ at time $0 \le s \le T$, an investment strategy $(L, M, c)$ is said to be admissible if the corresponding wealth processes $(X_t, Y_t)$ determined by \eqref{equ evolution1}-\eqref{equ evolution2} satisfy $(X_t, Y_t) \in \mathscr{S}$ for all $t \in [s, T]$. We denote by $\mathcal{A}_s(x, y)$ the set of all admissible strategies starting from $(x, y)$ at time $s$.

The investor aims to choose an admissible strategy to maximize the expected utility of discounted cumulative consumption and terminal wealth:
\begin{equation}\label{equ investors problem}
	\sup_{(L, M, c) \in \mathcal{A}_0(x, y)} 
	E_0^{x, y}\left[ \int_0^T e^{-\beta t} U(c_t) \, dt + e^{-\beta T} U(W_T) \right],
\end{equation}
subject to \eqref{equ evolution1}-\eqref{equ evolution2}, where $\beta > 0$ is the discount factor, $E_t^{x, y}$ denotes the conditional expectation given $(X_t, Y_t) = (x, y)$, and $U(\cdot)$ is the investor’s utility function.

We focus on the logarithmic utility,
\[
	U(c) = \log c,
\]
although our method also applies to general CRRA utilities $U(c) = c^\gamma / \gamma$ for $\gamma < 1$, $\gamma \neq 0$, in the case without consumption.

\subsection{Value function and HJB equation}\label{sec:value HJB}

Define the value function by
\begin{align}\label{equ value}
	\varphi(x, y, t)
	= \sup_{(L, M, c) \in \mathcal{A}_t(x, y)}
	E_t^{x, y}\left[
		\int_t^T e^{-\beta (s-t)} U(c_s)\, ds
		+ e^{-\beta (T-t)} U(W_T) 
	\right],
\end{align}
for $(x, y) \in \mathscr{S}$ and $0 \le t \le T$.  

The value function $\varphi(x, y, t)$ satisfies the Hamilton-Jacobi-Bellman (HJB) equation \citep[cf.][]{FlemingSoner2006, Shreve1994}:
\begin{equation}\label{varphi log}
	\min \left\{
		-\varphi_t - \mathcal{L} \varphi + \mathcal{A} \varphi,
		-(1-\mu)\varphi_x + \varphi_y,
		(1+\lambda)\varphi_x - \varphi_y
	\right\} = 0,
	\qquad (x, y) \in \mathscr{S},\ t \in [0, T),
\end{equation}
with terminal condition
\begin{align}\label{equ termi}
	\varphi(x, y, T) = U(x + (1-\mu)y),
\end{align}
where
\[
	\mathcal{L} \varphi
	= \frac{1}{2}\sigma^{2} y^{2} \varphi_{yy}
	+ \alpha y \varphi_y
	+ r x \varphi_x
	- \beta \varphi,
	\qquad
	\mathcal{A}\varphi = 1 + \ln \varphi_x.
\]

Due to the homogeneity of logarithmic utility, \eqref{equ value} implies
\begin{align}\label{equ homo vphi}
	\varphi(x, y, t)
	= \varphi\!\left(\frac{x}{y}, 1, t\right)
	+ g(t)\, \log y,
\end{align}
where $g(t) = \frac{1 - e^{-\beta (T-t)}}{\beta} + e^{-\beta(T-t)}>0$.
Define\footnote{See \citet{Dai2009Consumption}. For CRRA utility $U(c)=c^\gamma/\gamma$, $\gamma <1$, $\gamma\neq 0$, one defines $w(x,t)=\frac{1}{\gamma}\log(\gamma \varphi(x,1,t))$.}
\begin{equation}\label{key0}
	w(x, t) = \varphi(x, 1, t) / g(t),
\end{equation}
so that $\varphi(x, y, t) = g(t)\big(w(\frac{x}{y}, t) + \log y\big)$.%
\footnote{Introducing $w$ transforms the problem into one involving the bond-to-stock ratio $x/y$, instead of the fraction of wealth invested in stock.}

The function $w$ can be interpreted as the value function restricted to the line $y=1$. Substituting the representation into \eqref{varphi log} yields the reduced problem:
\begin{eqnarray}\label{equ w pde}
	\left\{
	\begin{array}{ll}
		\min \left\{
			-w_t - \mathcal{L}_0 w + \mathcal{A}_0 w,\;
			\frac{1}{x+1-\mu} - w_x,\;
			w_x - \frac{1}{x+1+\lambda}
		\right\} = 0,
		& (x,t) \in \Omega_T, \\[1.2ex]
		w(x, T) = \log(x + 1 - \mu), &
	\end{array}
	\right.
\end{eqnarray}
where $\Omega_T = (-(1-\mu), +\infty) \times [0, T)$, and
\begin{equation}\label{oper}
	\mathcal{L}_0 w
	= \frac{1}{2}\sigma^2 x^2 w_{xx}
	- (\alpha - r - \sigma^2)x w_x
	+ \left(\alpha - \frac{1}{2}\sigma^2\right),
\end{equation}
\[
	\mathcal{A}_0 w
	= \frac{1}{g(t)}\left(1 + \ln g(t) + w + \ln w_x\right).
\]

Define $v(x,t) : =  w_x(x,t)$. It has been shown that  
\[
	v(x, t) \in W^{2,1}_{p, \mathrm{loc}}(\Omega_T \setminus \{|x|<\delta\}) \cap C(\Omega_T),
	\qquad \forall\, \delta >0,\ 1 \le p < \infty,
\]
and $v$ is the unique weak solution to the following double-obstacle problem \citep[see][]{Dai2009Consumption}:
\begin{equation}\label{double ob pro}
\begin{cases}
	-v_t - \mathcal{L}_1 v + \mathcal{A}_1 v = 0,
	& \text{if }  \frac{1}{x+1+\lambda} < v(x,t) < \frac{1}{x+1-\mu}, \\[0.8ex]
	-v_t - \mathcal{L}_1 v + \mathcal{A}_1 v \le 0,
	& \text{if } v(x,t)=\frac{1}{x+1-\mu}, \\[0.8ex]
	-v_t - \mathcal{L}_1 v + \mathcal{A}_1 v \ge 0,
	& \text{if } v(x,t)= \frac{1}{x+1+\lambda}, \\[0.8ex]
	v(x, T) =  \frac{1}{x+1-\mu},
	& x > -(1-\mu),
\end{cases}
\end{equation}
where
\[
	\mathcal{L}_1 v
	= \frac{1}{2}\sigma^{2} x^{2} v_{xx}
	- (\alpha - r - 2\sigma^{2})x v_x
	- (\alpha - r - \sigma^{2}) v,
\qquad
	\mathcal{A}_1 v
	= \frac{1}{g(t)}\left( v + \frac{v_x}{v} \right).
\]

Problem \eqref{double ob pro} shows that the original singular control problem is equivalent to an optimal stopping formulation.

\subsection{Trading and no-trading regions}

Define the selling, buying, and no-trading regions by
\[
\begin{aligned}
	\mathbf{SR} &= \left\{(x, t) \in \Omega_T : w_x(x, t) = \frac{1}{x+1-\mu} \right\}, \\
	\mathbf{BR} &= \left\{(x, t) \in \Omega_T : w_x(x, t) = \frac{1}{x+1+\lambda} \right\}, \\
	\mathbf{NT} &= \left\{(x, t) \in \Omega_T : \frac{1}{x+1+\lambda} < w_x(x, t) < \frac{1}{x+1-\mu} \right\}.
\end{aligned}
\]

As shown in \citet{Dai2009Consumption, DaiYang2016, DAI20091445}, there exist two functions $x_s(t)$ and $x_b(t)$ for $0 \le t \le T$, satisfying 
\[
	-1 + \mu < x_s(t) < x_b(t) \le +\infty,
\]
such that
\[
\begin{aligned}
	\mathbf{SR} &= \{(x, t) \in \Omega_T : x \le x_s(t)\}, \qquad
	\mathbf{BR} = \{(x, t) \in \Omega_T : x \ge x_b(t)\}, \\
	\mathbf{NT} &= \{(x, t) \in \Omega_T : x_s(t) < x < x_b(t)\}.
\end{aligned}
\]
The functions $x_s(t)$ and $x_b(t)$ are referred to as the sell and buy boundaries, respectively. It is also known \citep[see][]{DAI20091445, Dai2009Consumption} that
\begin{align}\label{mm}
	x_b(t) > x_s(t) \ge 0 \qquad \text{when } \alpha - r \le \sigma^2 .
\end{align}

\begin{remark}
\qsj{Our free boundaries are defined in terms of the bond-to-stock ratio $x$, which corresponds to a stock proportion of $\frac{1}{x+1}$ in total wealth. A larger value of $x$ therefore represents a smaller proportion invested in the stock. In particular, $x=0$ (resp. $x=\infty$) means all wealth is invested in the stock (resp. no stock is held). Negative values of $x$ correspond to leveraged positions. The state variable must satisfy $x \ge -(1-\mu)$, since values below this threshold imply a negative wealth level.}
\end{remark}

\section{\qsj{Impact of transaction cost rates}}\label{Sec mono cost}

\qsj{In this section, we theoretically examine how the transaction cost rates $\mu$ and $\lambda$ affect the trading boundaries. We first show that it is the cost-adjusted trading boundaries—rather than the trading boundaries themselves—that exhibit monotonicity with respect to these rates. We then establish that when transaction costs are sufficiently high, the sell boundary becomes independent of the purchase cost rate $\lambda$. }

\subsection{\qsj{Monotonicity in $\lambda$ and $\mu$}}

The following theorem summarizes our main result.
\begin{theorem}\label{thm: transaction con monotone}
Let $x_s(t; \lambda,\mu)$ and $x_b(t; \lambda,\mu)$ denote the sell and buy boundaries of the finite-horizon problem \eqref{equ investors problem} with transaction cost rates $(\lambda,\mu)$. Then:

\noindent (i) The sell boundary $x_s(t;\lambda,\mu)$ is monotonically decreasing in the purchase cost $\lambda$, and the buy boundary $x_b(t;\lambda,\mu)$ is monotonically increasing in the sale cost $\mu$.

\noindent (ii) The cost-adjusted sell boundary $\frac{x_s(t;\lambda,\mu)}{1-\mu}$ is monotonically decreasing in $\mu$, and the cost-adjusted buy boundary $\frac{x_b(t;\lambda,\mu)}{1+\lambda}$ is monotonically increasing in $\lambda$.
\end{theorem}
We refer to $\frac{x_s(t;\lambda,\mu)}{1-\mu}$ and $\frac{x_b(t;\lambda,\mu)}{1+\lambda}$ as the cost-adjusted sell and buy boundaries. Note that the sell boundary must be adjusted using the sale cost, and the buy boundary must be adjusted using the purchase cost. 
These adjusted boundaries may be interpreted as the critical values of a modified bond-to-stock ratio at which trading occurs, where the stock position is evaluated at the ask or bid price as appropriate. Because $\frac{x}{1-\mu} = \frac{X_t}{(1-\mu)Y_t}$ and $\frac{x}{1+\lambda} = \frac{X_t}{(1+\lambda)Y_t}$ correspond to the bond-to-stock ratios obtained by valuing the stock position at the bid and ask prices, respectively.

Because the investor must liquidate all stock holdings at time $T$, the relevant measure of stock value at time $t$ is the post-liquidation value $(1-\mu)Y_t$, rather than the pre-liquidation holding $Y_t$. Consequently, on the sell boundary an efficient portfolio allocation requires
\[
\frac{X_t}{(1-\mu)Y_t} = \frac{x_s(t)}{1-\mu}
\]
to be below the Merton line $x_M$. This is consistent with our result that the cost-adjusted sell boundary $x_s(t)/(1-\mu)$ is decreasing in the transaction cost rates.

Mathematically, unlike the original free boundaries, the cost-adjusted boundaries depend on the transaction cost rates $\mu$ and $\lambda$ only through the ratio $\frac{1+\lambda}{1-\mu}$; see the proof of Theorem \ref{thm: transaction con monotone} in Section \ref{sec proof thm1}. This structural property facilitates the monotonicity analysis.

\begin{remark}\label{rmk xb = infty}
The monotonicity may not be strict. Theorem 4.6 of \citet{Dai2009Consumption} proves that 
\[
x_b(t;\lambda,\mu) = +\infty \quad \text{if} \quad 
t \ge T - \frac{1}{\alpha - r} \log\!\left(\frac{1+\lambda}{1-\mu}\right),
\]
meaning that the investor will no longer buy any stock after this time. Moreover, once no further purchases occur, increasing the purchase cost rate $\lambda$ does not affect the sell boundary $x_s(t)$.
\end{remark}

From Theorem \ref{thm: transaction con monotone}, we immediately obtain the following corollary.

\begin{corollary}\label{c1}
The cost-adjusted sell boundary $\frac{x_s(t;\lambda,\mu)}{1-\mu}$ is monotonically decreasing in both $\lambda$ and $\mu$, and the cost-adjusted buy boundary $\frac{x_b(t;\lambda,\mu)}{1+\lambda}$ is monotonically increasing in both $\lambda$ and $\mu$.
\end{corollary}

If $\alpha - r \le \sigma^2$, Theorem \ref{thm: transaction con monotone} further implies that the no-trading region expands as the transaction cost rates increase.

\begin{corollary}\label{corr mono cost}
Assume $\alpha - r \le \sigma^2$. Then $x_s(t;\lambda,\mu)$ is monotonically decreasing in both $\lambda$ and $\mu$, while $x_b(t;\lambda,\mu)$ is monotonically increasing in both $\lambda$ and $\mu$.
\end{corollary}

\noindent\textbf{Proof of Corollary \ref{corr mono cost}.}
By part (i) of Theorem \ref{thm: transaction con monotone}, it suffices to prove that $x_s(t;\lambda,\mu)$ is decreasing in $\mu$ and $x_b(t;\lambda,\mu)$ is increasing in $\lambda$. We prove the former; the latter follows analogously.  

Part (ii) of Theorem \ref{thm: transaction con monotone} gives
\begin{align}\label{eq1}
\frac{x_s(t;\lambda,\mu_1)}{1-\mu_1} \ge \frac{x_s(t;\lambda,\mu_2)}{1-\mu_2}
\qquad \text{if } \mu_2 > \mu_1.
\end{align}
Under $\alpha - r \le \sigma^2$, relation \eqref{mm} ensures $x_s(t; \lambda,\mu_2) \ge 0$, which yields
\begin{align}\label{eq2}
\frac{x_s(t;\lambda,\mu_2)}{1-\mu_2} \ge \frac{x_s(t;\lambda,\mu_2)}{1-\mu_1}.
\end{align}
Combining \eqref{eq1} and \eqref{eq2} proves the result. \qed

\begin{remark}
As shown in \citet{DAI20091445} and \citet{Dai2009Consumption},
\[
\lim_{t\to T-} x_s(t;\lambda,\mu) = (1-\mu)x_M,
\qquad
x_M := -1 + \frac{\sigma^2}{\alpha - r},
\]
where $x_M$ is the Merton line.\footnote{\qsj{This corresponds to a stock fraction $\frac{\alpha - r}{\sigma^2}$.}}  
This demonstrates that the assumption $\alpha - r \le \sigma^2$ is necessary for part (ii) of Corollary \ref{corr mono cost}; otherwise, $x_M < 0$ and the monotonicity fails as $t \to T$.\end{remark}

Before proving Theorem \ref{thm: transaction con monotone}, we highlight its financial implications.  
When $\lambda = \mu = 0$, the trading boundaries coincide with the Merton line:
\[
x_s(t;0,0) = x_b(t;0,0) = x_M.
\]
When $x_M \ge 0$, it must lie inside the no-trading region, i.e.,
\begin{align}\label{equ con notrading region}
x_s(t;\lambda,\mu) \le x_M \le x_b(t;\lambda,\mu),
\end{align}
a conclusion also obtainable from Corollary \ref{corr mono cost}.  
However, Corollary \ref{c1} yields the following stronger result.\footnote{By \eqref{mm}, inequality \eqref{equ con xm xs xb} implies \eqref{equ con notrading region} when $x_M \ge 0$.}

\begin{corollary}
The Merton line always lies between the two cost-adjusted trading boundaries:
\begin{align}\label{equ con xm xs xb}
\frac{x_s(t;\lambda,\mu)}{1-\mu}
\ \le\
x_M
\ \le\
\frac{x_b(t;\lambda,\mu)}{1+\lambda}.
\end{align}
\end{corollary}

When $x_M < 0$, the Merton line may or may not lie inside the no-trading region, depending on the magnitude of transaction costs and the time to maturity (see \citep{Shreve1994, Hobson2019}).  
However, inequality \eqref{equ con xm xs xb} remains valid regardless of the sign of $x_M$.

For fixed $\lambda$ and $\mu$, Theorem 4.6 of \citet{Dai2009Consumption} shows that $x_b(t;\lambda,\mu) = +\infty$ for $t$ sufficiently close to $T$. When $\alpha - r > \sigma^2$, $x_b(t;\lambda,\mu)$ may become negative for some $t$. We therefore define
\[
\tau_1 := \sup\{\, t \le T : x_b(t;\lambda,\mu) \le 0 \,\},
\]
the last time at which the buy boundary is non-positive.  We refer $\tau_1 = -\infty$ to the case that $x_b(t;\lambda,\mu) > 0$ for all $t\leq T$. 

Proposition 4.1 of \citet{Dai2009Consumption} shows that in the absence of consumption,
\[
\tau_1 = T - \frac{1}{\alpha - r - \sigma^2}\log\!\left(\frac{1+\lambda}{1-\mu}\right),
\]
while \citet{Hobson2019} indicate that $x^*_b(\lambda, \mu)>0$ when $\frac{1+\lambda}{1-\mu}$ is sufficiently large, where $x^*_b(\lambda, \mu)$ is the purchase boundary of an infinite-horizon consumption-investment problem. Although a complete characterization of when $\tau_1 = -\infty$ is unavailable due to the non-monotonicity of the free boundaries in time (see \citep{DaiZhong2008}), Theorem \ref{thm: transaction con monotone} implies the following result.
\begin{corollary}\label{coro xb sign}
For any $\alpha - r - \sigma^2 > 0$,\footnote{When $\alpha - r - \sigma^2 < 0$, \eqref{mm} shows that $\tau_1 = -\infty$.} the quantity $\tau_1 \in [-\infty,\, T)$ is monotonically decreasing in $\lambda$ and $\mu$.
\end{corollary}

\subsection{{Sell boundary when $\theta := \frac{1+\lambda}{1-\mu}$ is large}}

According to Corollary 5.4 of \citet{Hobson2019}, in the infinite-horizon setting, when $\alpha - r - \sigma^2 > 0$, there exists a cutoff value\footnote{\citet{Hobson2019} use the notation $\bar{\xi}$ for our $\bar{\theta}$, and the explicit definition is given therein.} $\bar{\theta}$ such that the optimal buy boundary $x_b^* > 0$ if and only if $\theta \ge \bar{\theta}$. Moreover, in this case, the optimal sell boundary $x_s^*$ is independent of the purchase cost rate $\lambda$, where $x_s^*$ and $x_b^*$ denote the stationary sell and buy boundaries in the infinite-horizon problem.

\qsj{Intuitively, when transaction costs are high, if the investor starts with positive cash, it is suboptimal to purchase stock to reach a leveraged position. Instead, the investor prefers to buy fewer shares and retain cash for future consumption, which corresponds to $x_b^* > 0$. In contrast, if the investor starts with negative cash or consumes enough that the cash balance becomes negative, then the optimal policy is simply to borrow for consumption and sell stock upon reaching the sell boundary. In either case, the investor will not purchase additional stock, and therefore the purchase cost rate $\lambda$ will not affect the value function for $x < 0$ or the location of the sell boundary. This intuition is also reflected in the analysis of \citet{Hobson2019}.}

The following theorem confirms that the same phenomenon occurs in our finite-horizon problem, driven by the same intuition.

\begin{theorem}\label{thm large theta}
When $\alpha - r - \sigma^2 > 0$, $\theta \ge \bar{\theta}$, and the discount rate $\beta \le 1$, we have $x_s(t) < 0 < x_b(t)$, and the sell boundary $\{x_s(t)\}_{0 \le t \le T}$ is independent of the purchase cost rate $\lambda$.
\end{theorem}

\begin{remark}
We cannot expect the buy boundary $x_b(t)$ to be independent of the sale cost rate $\mu$, because the terminal condition depends on the wealth after liquidation, which necessarily incorporates $\mu$.
\end{remark}

\subsection{Proof of Theorem \ref{thm: transaction con monotone}}\label{sec proof thm1}
Let us first consider an alternative formulation that merges the purchase and sale costs. Denote $\hat{S}_{t}:=(1-\mu) S_{t}$, which can be regarded as the bid price of the stock, and let $\hat{Y}_{t}:=(1-\mu) Y_{t}$ be the dollar amount invested in the stock in terms of this bid price. We then rewrite \eqref{equ evolution1}-\eqref{equ evolution2} as
\begin{equation}\label{reformulation dynamic}
\left\{
\begin{array}{l}
	d X_{t}=(r X_{t}-c_t)\, d t - \theta\, d \hat{L}_{t} + d \hat{M}_{t}, \\
	d \hat{Y}_{t}=\alpha \hat{Y}_{t}\, d t + \sigma \hat{Y}_{t}\, d B_{t} + d \hat{L}_{t} - d \hat{M}_{t},
\end{array}
\right.
\end{equation}
where $\hat{L}_t := (1-\mu) L_t$ and $\hat{M}_t := (1-\mu) M_t$. Consequently, the wealth process becomes $W_t = X_t + \hat{Y}_t$. The investor’s problem \eqref{equ investors problem} is then rewritten as
\begin{align}\label{reformulation target}
\sup_{(\hat{L}, \hat{M}, c)} 
E_0^{x, \hat{y}}
\left[
\int_0^T e^{-\beta t} U(c_t)\, dt + e^{-\beta T} U(W_T)
\right],
\qquad
\text{s.t. } W_t \ge 0,\ \forall\, t \ge 0,
\end{align}
which is equivalent to the original model with zero sale cost and purchase cost $\theta - 1$. Denote by $\hat{\varphi}(x, \hat{y}, t)$ the corresponding value function with initial condition $X_t = x$ and $\hat{Y}_t = \hat{y}$. It then follows immediately that
\[
\hat{\varphi}(x, \hat{y}, t)
=
\varphi\!\left(x, \frac{\hat{y}}{1-\mu}, t\right).
\]

By homogeneity in \eqref{equ homo vphi}, we define
\begin{eqnarray}\label{nhom}
\hat{w}(x, t)=\frac{\hat{\varphi}(x, 1, t)}{g(t)}, 
\qquad 
\hat{x}=\frac{x}{1-\mu}.
\end{eqnarray}
It follows that
\begin{eqnarray*}
\hat{w}(\hat{x}, t)
&=&
\frac{\hat{\varphi}(\hat{x}, 1, t)}{g(t)}
=
\frac{\varphi\!\left(\hat{x}, \frac{1}{1-\mu}, t\right)}{g(t)}
=
\frac{\varphi(x, 1, t)}{g(t)} - \log(1-\mu)
\\
&=&
w(x, t) - \log(1-\mu).
\end{eqnarray*}
Using the HJB equation \eqref{equ w pde} satisfied by $w(x,t)$, we obtain the corresponding HJB equation for $\hat{w}(\hat{x}, t)$:
\begin{eqnarray}
\left\{
\begin{array}{ll}
\min\!\left\{
- \hat{w}_t - \mathcal{L}_{0}\hat{w} + \mathcal{A}_{0}\hat{w},\;
\frac{1}{\hat{x}+1} - \hat{w}_{\hat{x}},\;
\hat{w}_{\hat{x}} - \frac{1}{\hat{x}+\theta}
\right\}=0,
& (\hat{x},t)\in \hat{\Omega}_T,
\\[3pt]
\hat{w}(\hat{x}, T)=\log(\hat{x}+1), &
\end{array}
\right.
\label{equ hatobstacleproblem}
\end{eqnarray}
where $\hat{\Omega}_T = (-1, +\infty)\times [0, T)$.

Equation \eqref{equ hatobstacleproblem} induces two free boundaries: the sell and buy boundaries, denoted by $\hat{x}_s(t)$ and $\hat{x}_b(t)$, respectively. It is immediate that
\begin{align}
\hat{x}_s(t)=\frac{x_s(t)}{1-\mu}.
\end{align}

Similarly, if we consider the ask price $\overline{S}_t := (1+\lambda) S_t$ and define $\overline{Y}_t := (1+\lambda) Y_t$, then we can introduce
\begin{align}
\overline{x}_b(t)=\frac{x_b(t)}{1+\lambda},
\end{align}
which represents the buy boundary associated with zero purchase cost and sale cost $1 - \frac{1}{\theta}$.

We only need to show that $\hat{x}_s(t)$ is monotonically decreasing with respect to $\theta$ and that $\bar{x}_b(t)$ is monotonically increasing with respect to $\theta$. These two properties immediately yield parts (i) and (ii) of Theorem \ref{thm: transaction con monotone}. In what follows, we focus on the HJB equation \eqref{equ hatobstacleproblem} to establish the monotonicity of $\hat{x}_s$ in $\theta$, namely,
\begin{eqnarray}\label{kth}
\hat{x}_s(t;\theta_1)\geq \hat{x}_s(t;\theta_2)
\qquad \text{whenever } \theta_1<\theta_2,
\end{eqnarray}
where $\hat{x}_s(t;\theta_i)$ denotes the sell boundary associated with $\theta=\theta_i$, $i=1,2$.  
The monotonicity of $\bar{x}_b$ is proved in the same manner.

Analogous to the double obstacle problem for $w$, we next study its counterpart for $\hat{w}$. For simplicity of notation, we use the state variable $x$ in place of $\hat{x}$. Define 
\[
\hat{v}(x,t) \equiv \hat{w}_x(x,t).
\]
It is known that $\hat{v}(x,t)\in W^{2,1}_{p,\mathrm{loc}}(\hat{\Omega}_T \setminus \{|x|<\delta\}) \cap C(\hat{\Omega}_T)$ for any $\delta>0$ and any $1\le p<\infty$, and that $\hat{v}$ is the unique weak solution to the following double obstacle problem:
\begin{equation}
\left\{
\begin{array}{l}
-\hat{v}_t - \mathcal{L}_1 \hat{v} + \mathcal{A}_1 \hat{v} = 0 
\quad \text{if }  \frac{1}{x+\theta} < \hat{v} < \frac{1}{x+1}, \\[4pt]
-\hat{v}_t - \mathcal{L}_1 \hat{v} + \mathcal{A}_1 \hat{v} \le 0 
\quad \text{if } \hat{v}=\frac{1}{x+1}, \\[4pt]
-\hat{v}_t - \mathcal{L}_1 \hat{v} + \mathcal{A}_1 \hat{v} \ge 0 
\quad \text{if }  \hat{v}=\frac{1}{x+\theta}, \\[4pt]
\hat{v}(x,T) = \frac{1}{x+1},
\end{array}
\right.
\label{equ obstacleproblem}
\end{equation}
posed on $\hat{\Omega}_T$.

The double obstacle problem \eqref{equ obstacleproblem} gives rise to two free boundaries corresponding to the buy and sell boundaries $\hat{x}_b(t)$ and $\hat{x}_s(t)$ defined above. The function $\hat{v}(x,t)$ also characterizes the trading and no-trading regions:
\[
\begin{aligned}
\mathbf{S R} 
&= \left\{(x,t)\in \hat{\Omega}_T : \hat{v}(x,t)=\frac{1}{x+1}\right\}
 = \left\{(x,t)\in \hat{\Omega}_T : x\le \hat{x}_s(t)\right\}, \\[4pt]
\mathbf{B R} 
&= \left\{(x,t)\in \hat{\Omega}_T : \hat{v}(x,t)=\frac{1}{x+\theta}\right\}
 = \left\{(x,t)\in \hat{\Omega}_T : x\ge \hat{x}_b(t)\right\}, \\[4pt]
\mathbf{N T} 
&= \left\{(x,t)\in \hat{\Omega}_T : \frac{1}{x+\theta}<\hat{v}(x,t)<\frac{1}{x+1}\right\}
 = \left\{(x,t)\in \hat{\Omega}_T : \hat{x}_s(t)<x<\hat{x}_b(t)\right\}.
\end{aligned}
\]

Our proof relies on the following comparison principle for the double obstacle problem \eqref{equ obstacleproblem}.
\begin{lemma}\label{prop compare transa cost}
Let $\hat{v}(x,t;\theta)$ be the solution to the double obstacle problem \eqref{equ obstacleproblem}. If $\theta_1\le \theta_2$, then
\[
\hat{v}(x,t;\theta_1) \ge \hat{v}(x,t;\theta_2)
\qquad \text{in } \hat{\Omega}_T.
\]
\end{lemma}
The proof of this lemma is relegated to the Appendix \ref{sec proof lem1}.

We are now ready to prove Theorem \ref{thm: transaction con monotone}.

\medskip
\noindent{\bf Proof of part (i).}
We need to show that $x_s(t;\lambda_1,\mu)\ge x_s(t;\lambda_2,\mu)$ for all $t<T$ whenever $\lambda_1\le \lambda_2$. Suppose to the contrary that $x_s(t;\lambda_1,\mu)<x_s(t;\lambda_2,\mu)$ for some $t<T$. Then
\[
\hat{x}_s(t;\theta_1)<\hat{x}_s(t;\theta_2),
\]
and therefore
\[
\hat{v}(\hat{x}_s(t;\theta_2),t;\theta_1) 
< \frac{1}{\hat{x}_s(t;\theta_2)+1}
= \hat{v}(\hat{x}_s(t;\theta_2),t;\theta_2),
\]
which contradicts Lemma \ref{prop compare transa cost}.  
The argument for the buy boundary is analogous.

\medskip
\noindent{\bf Proof of part (ii).}
It suffices to prove the monotonicity of 
\[
\hat{x}_s(t;\theta)=\frac{x_s(t;\lambda,\mu)}{1-\mu}
\]
with respect to $\mu$. As in part (i), $\hat{x}_s(t;\theta)$ is decreasing in $\theta$, and since $\theta$ is increasing in $\mu$, the claim follows.

\begin{remark}
In the case $U(c)=\frac{c^\gamma}{\gamma}$ with consumption, the differential operator $\mathcal{A}_1$ in the double obstacle problem \eqref{equ obstacleproblem} depends on $\hat{w}(x,t)$ (see \cite{Dai2009Consumption}). Specifically,
\[
\mathcal{A}_1 \hat{v}(x,t)
=
e^{-\frac{\gamma}{1-\gamma}\hat{w}}
\left[
\hat{v}^{\,2-\frac{1}{1-\gamma}}
+\hat{v}^{-\frac{1}{1-\gamma}}\, \hat{v}_x
\right],
\qquad
\gamma\ne 0,\ \gamma<1.
\]
Hence the problem is not self-contained, and the associated comparison principle remains open. Nevertheless, we conjecture that Theorem \ref{thm: transaction con monotone} continues to hold for power utility with consumption, and we leave this question for future study.
\end{remark}

\section{\qsj{Numerical results}}\label{sec numerical}
In this section, we provide numerical results to illustrate the behavior of the free boundaries. We focus on the more interesting case with a negative Merton line ($x_M<0$). Numerical results for the case with a positive Merton line are relegated to Appendix \ref{sec app posi merton}. 

\subsection{Trading boundaries with respect to time}
\begin{figure}[htbp]
\begin{center}
\centering
\includegraphics[width=0.6\textwidth]{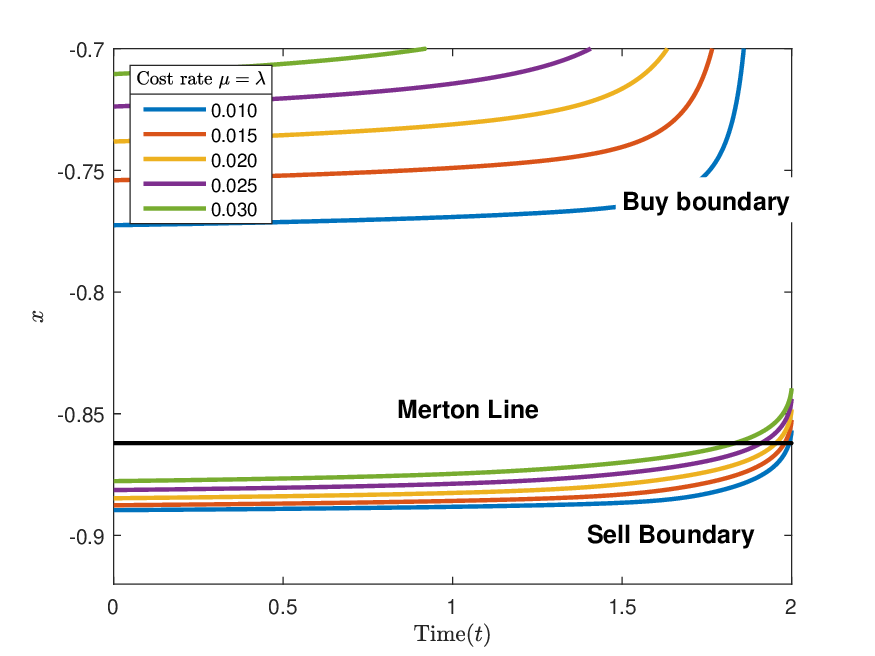}
\caption{\footnotesize Trading boundaries against time. Parameters: $T=2$, $\alpha = 0.3$, $r = 0.01$, and $\sigma = 0.2$. The corresponding Merton line is $x_M = -0.862$.}
\label{fig:tranding boundary with t}
\end{center}
\end{figure}

We first plot the trading boundaries under different transaction cost rates in the following Figure \ref{fig:tranding boundary with t}. The sell boundaries are not monotone decreasing in transaction cost rates. Instead, under our parameters, it is actually increasing in the time scope $t\in [0,2]$.  These observations further highlight Theorem \ref{thm: transaction con monotone} and Corollaries \ref{c1} and \ref{corr mono cost}. 

For fixed $\mu$, $\lambda$, the no-trading region is not monotone in time, a phenomenon similar to the findings of \citet{dai2024dynamic}. The reason here is that both trading boundaries are increasing in time in our current numerical parameters.    
 Moreover, as demonstrated in \citet{DaiZhong2008}, the trading boundaries themselves may fail to be monotone, which further destroys the monotonicity of the no-trading region in time.

\subsection{Trading boundaries with respect to transaction cost rates}\label{sec tb cost}

In this subsection, motivated by Figure \ref{fig:tranding boundary with t}, we fix time $t$ and examine how the transaction costs affect the free boundaries. 
\begin{figure}[htbp]
\begin{center}
\centering
\includegraphics[width=1\textwidth]{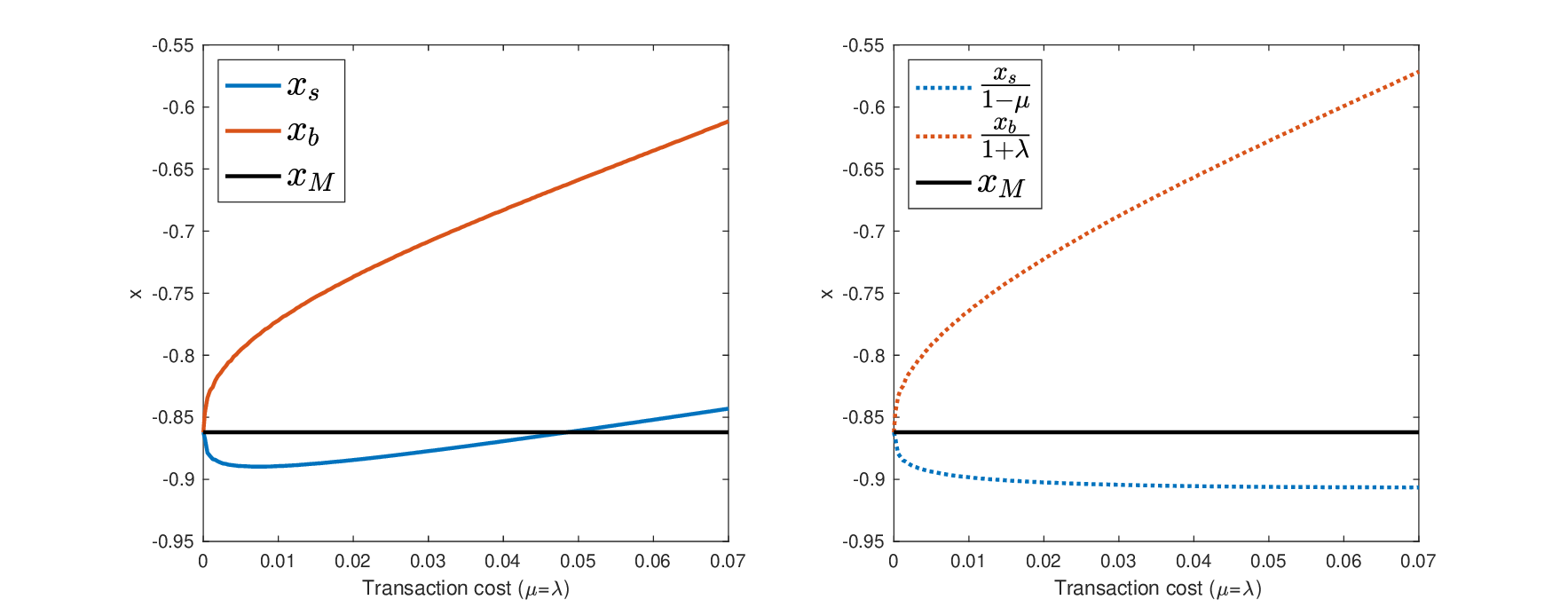}
\caption{\footnotesize Trading boundaries (left) and cost-adjusted trading boundaries (right) for $\mu=\lambda$. Default parameters: $T=2$, $t=0.25$, $\alpha=0.3$, $r=0.01$, and $\sigma=0.2$. The corresponding Merton line is $x_M=-0.862$.}
\label{fig:noFixedNegative}
\end{center}
\end{figure}

In Figure \ref{fig:noFixedNegative}, we set $\mu=\lambda$ and plot the trading boundaries and the cost-adjusted trading boundaries. The left panel shows that, when transaction costs are small, the sell boundary (blue solid line) lies below the Merton line (black line), whereas for larger transaction costs the sell boundary lies above the Merton line. The right panel illustrates that the Merton line always lies between the two cost-adjusted trading boundaries. 
\begin{figure}[htbp]
\begin{center}
\centering
\includegraphics[width=1\textwidth]{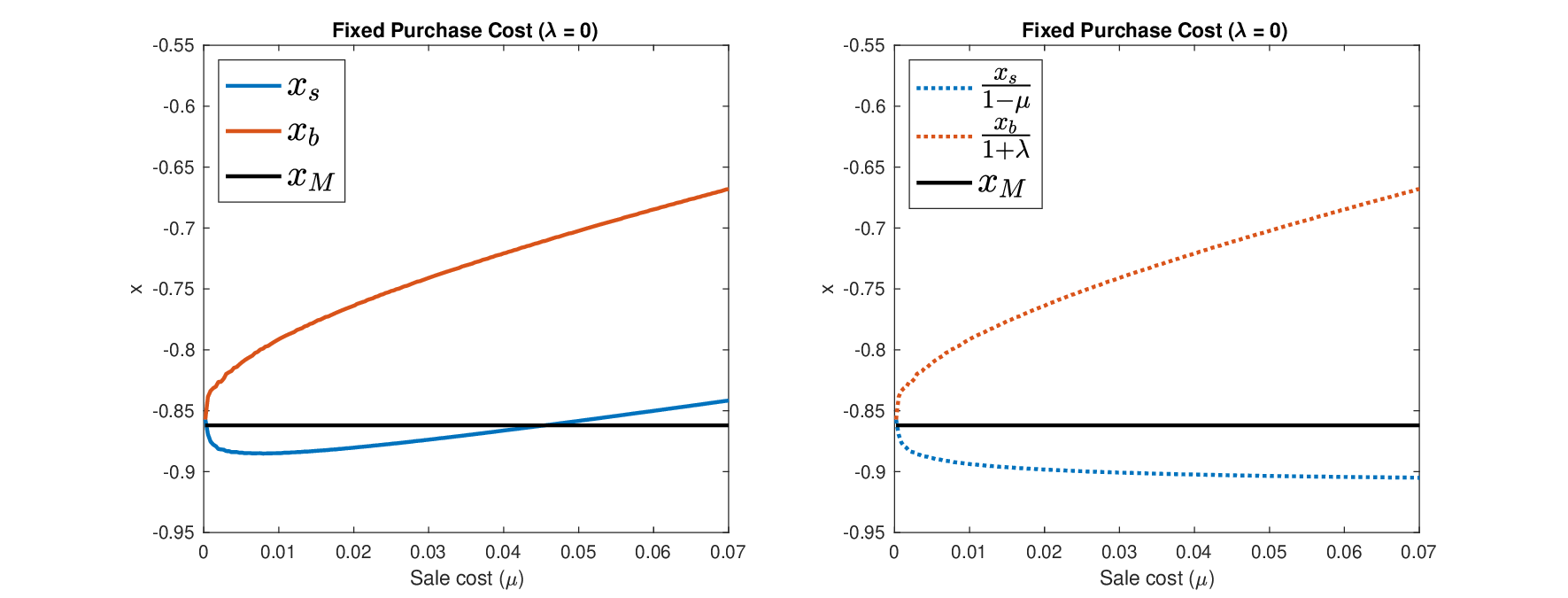}
\caption{\footnotesize Trading boundaries (left) and cost-adjusted trading boundaries (right) against the sale cost rate $\mu$. Default parameters: $T=2$, $t=0.25$, $\alpha=0.3$, $r=0.01$, and $\sigma=0.2$. The corresponding Merton line is $x_M=-0.862$.}
\label{fig:buyFixedNegative}
\end{center}
\end{figure}

\begin{figure}[htbp]
\begin{center}
\centering
\includegraphics[width=1\textwidth]{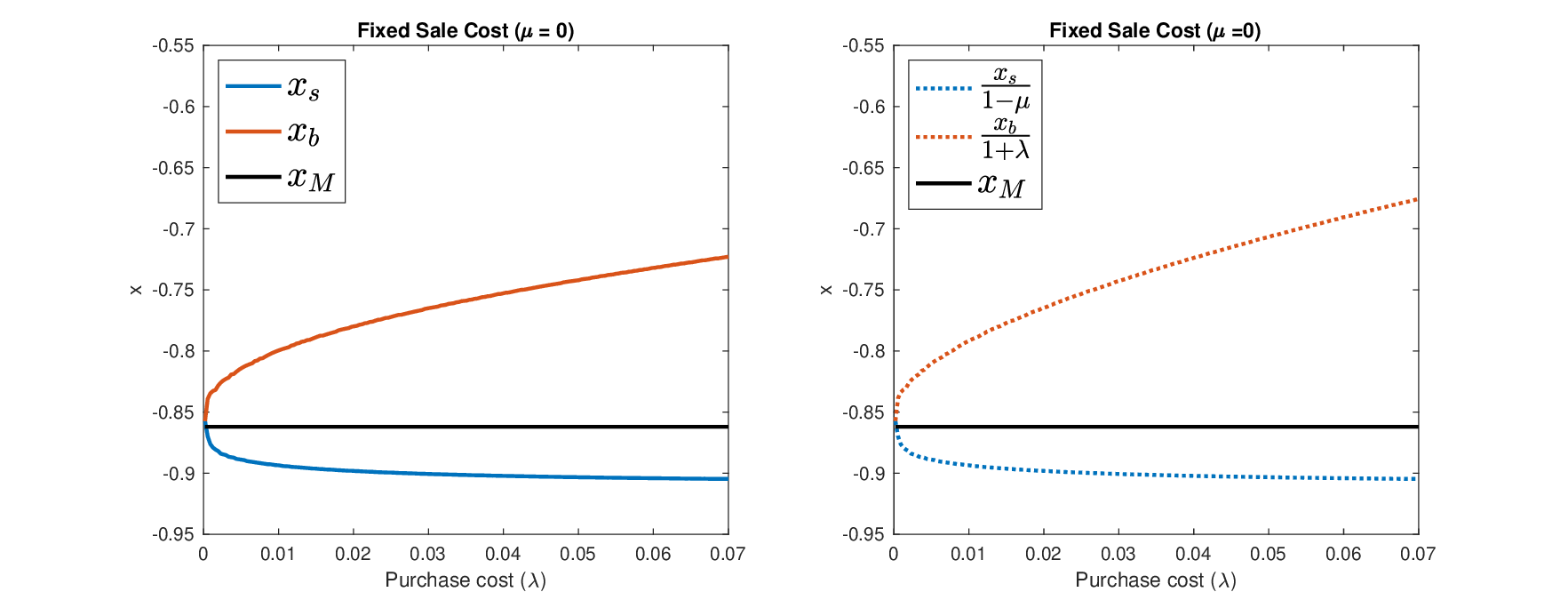}
\caption{\footnotesize Trading boundaries (left) and cost-adjusted trading boundaries (right) against the purchase cost rate $\lambda$. Default parameters: $T=2$, $t=0.25$, $\alpha=0.3$, $r=0.01$, and $\sigma=0.2$. The corresponding Merton line is $x_M=-0.862$.}
\label{fig:sellFixedNegative}
\end{center}
\end{figure}

\qsj{A similar pattern appears when fixing $\lambda$ and varying $\mu$, as shown in Figure \ref{fig:buyFixedNegative}. In contrast, Figure \ref{fig:sellFixedNegative} shows that the Merton line lies within the no-trading region when the sale cost $\mu$ is fixed and the purchase cost $\lambda$ is varied. This follows from the fact that the sell boundary is monotone decreasing in $\lambda$ (see Theorem \ref{thm: transaction con monotone}).
}

In Figure \ref{fig:tranding boundary with t}, we observe that the sell boundary may be higher than the Merton line $x_M$ when the time to maturity is short, which implies that the Merton line lies outside the no-trading region. In the infinite-horizon setting, such a phenomenon arises only when the transaction cost rates are sufficiently large (see \cite{Hobson2019}, \cite{janevcek2004asymptotic}, and \cite{Shreve1994}). In contrast, for the finite-horizon case, small transaction costs cannot guarantee that the Merton line always lies within the no-trading region, as \citet{Dai2009Consumption} prove that
\begin{align}\label{equ xs T}
x_s(T) = (1-\mu)x_M,
\end{align}
which exceeds $x_M$ whenever $x_M<0$.

Motivated by Figure \ref{fig:buyFixedNegative}, for fixed $t$ and $\lambda=0$, we define the critical value
\[
\mu^*(t) := \max\bigg\{\hat{\mu}\ge 0:\ x_s(t)\le x_M \text{ for all }\mu\in[0,\hat{\mu}]\bigg\}.
\]

\begin{figure}[htbp]
\begin{center}
\centering
\includegraphics[width=0.6\textwidth]{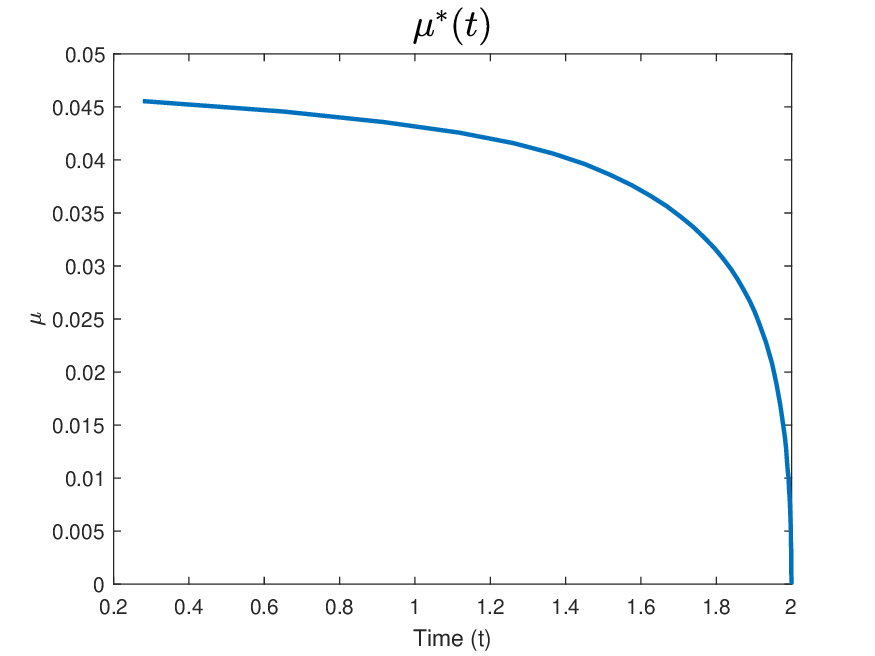}
\caption{\footnotesize The critical value $\mu^*(t)$. Default parameters: $T=2$, $\alpha=0.3$, $r=0.01$, and $\sigma=0.2$.}
\label{fig:t_muAccrossMerton}
\end{center}
\end{figure}

Figure \ref{fig:t_muAccrossMerton} plots $\mu^*(t)$. We see that $\mu^*(T)=0$ due to the boundary condition \eqref{equ xs T}. For any $t<T$, when transaction costs are sufficiently small, the Merton line lies above the sell boundary. Moreover, in our numerical experiments, we did not observe any instance where $x_b(t)$ falls below the Merton line. Intuitively, if the investor buys too many stocks, then (i) transaction costs incurred are higher and (ii) the risk exposure becomes inefficient (always exceeding the Merton strategy), making such a policy suboptimal.
Thus, for $\lambda= 0$, when $\mu\leq \mu^*(t)$, the Merton line is contained in the no-trading region at time $t$; as $t\to T$, this cutoff $\mu^*(t)$ decreases to $0$.

\section{Conclusion}\label{sec conclusion}
In this paper we investigate how the optimal trading boundary varies with transaction cost rates. Extant literature shows that when the Merton line is negative, it may fall outside the no-trading region, and consequently the trading boundary may not be monotone in the transaction costs. In contrast, we find that although the trading boundary itself may not be monotone, the cost-adjusted trading boundary {is} monotone in the transaction cost rates. As a result, the Merton line always lies between the cost-adjusted buy and sell boundaries.

\citet{Hobson2019} prove this result for the infinite-horizon case by exploiting a semi-closed-form value function for the one-dimensional problem. For the finite-horizon case, the value function depends on both the state variable and time, and their approach no longer applies. Instead, we make use of an equivalent optimal stopping formulation (Problem~\eqref{equ obstacleproblem}) and establish the corresponding comparison principle. However, in the power-utility-with-consumption setting, the associated optimal stopping problem is not self-contained, which complicates the analysis of comparison principle. As a consequence, our proof is valid only for the no-consumption case or the logarithmic utility case. We believe the result continues to hold in a more general framework and leave this extension for future research.

These theoretical findings are further verified by numerical analysis. The results show that when transaction cost rates are sufficiently high, a negative Merton line (i.e., a leveraged position) may lie below the sell boundary, implying that it is outside the no-trading region. Numerically, we also identify a time-dependent cutoff for the transaction cost rates such that the Merton line remains within the no-trading region; this cutoff converges to zero as the time to maturity approaches zero.

\clearpage
 \counterwithout{figure}{section}
\counterwithout{equation}{section}
\begin{appendices}

\section{Proof of Lemma \ref{prop compare transa cost}}\label{sec proof lem1}

We prove by contradiction. Denote
\[
\mathcal{N}=\left\{(x,t)\in\hat{\Omega}_T \,\big|\, \hat{v}(x,t;\theta_1)<\hat{v}(x,t;\theta_2)\right\},
\]
and assume that $\mathcal{N}$ is a nonempty open set. Then we must have
\[
\hat{v}(x,t;\theta_1)<\frac{1}{x+1}, 
\qquad 
\hat{v}(x,t;\theta_2)>\frac{1}{x+\theta_2},
\qquad \text{in } \mathcal{N}.
\]
It follows that, in $\mathcal{N}$,
\begin{align}
    -\hat{v}_t(x,t;\theta_1) - \mathcal{L}_1 \hat{v}(x,t;\theta_1) + \mathcal{A}_1 \hat{v}(x,t;\theta_1) &\ge 0, \label{equ cost 1}\\
    -\hat{v}_t(x,t;\theta_2) - \mathcal{L}_1 \hat{v}(x,t;\theta_2) + \mathcal{A}_1 \hat{v}(x,t;\theta_2) &\le 0. \label{equ cost 2}
\end{align}

Let $u(x,t)=\hat{v}(x,t;\theta_1)-\hat{v}(x,t;\theta_2)$.  
From \eqref{equ cost 1}--\eqref{equ cost 2}, we obtain
\begin{align}
\begin{cases}
\displaystyle
    -u_t - \frac{1}{2}\sigma^2 x^2 u_{xx} 
    + (\alpha - r - 2\sigma^2)x\,u_x
    + (\alpha - r - \sigma^2)u \\[6pt]
\displaystyle \qquad\qquad
    + \frac{1}{g(t)}\left(
        u + \frac{u_x}{\hat{v}(\cdot;\theta_1)}
        - \frac{\hat{v}_x(\cdot;\theta_2)\,u}{\hat{v}(\cdot;\theta_1)\hat{v}(\cdot;\theta_2)}
    \right)
    \ge 0, & \text{in } \mathcal{N}, \\[6pt]
    u = 0, & \text{on } \partial_p \mathcal{N}, \label{equ:comp2}
\end{cases}
\end{align}
where $\partial_p \mathcal{N}$ denotes the parabolic boundary of $\mathcal{N}$.

By treating $\hat{v}$ and $\hat{v}_x$ as known functions, the above PDE is linear in $u$.  
Applying the maximum principle yields $u \ge 0$ in $\mathcal{N}$, i.e.,
\[
\hat{v}(x,t;\theta_1) \ge \hat{v}(x,t;\theta_2)\quad \text{in } \mathcal{N},
\]
which contradicts the definition of $\mathcal{N}$.

\section{Proof of Theorem \ref{thm large theta}}
We first prove that $x_s(t)< 0 < x_b(t)$, $\forall\ t \in [0, T]$.  According to  \citet{Dai2009Consumption}, we have $x_s(t) \leq x_s(T-) \leq (1-\mu) x_M <0 $, the left inequality holds true. 

To handle the right inequality, we consider the infinite-horizon investment problem
\begin{equation}\label{infinite horizon problem}
\sup _{(L, M, c) \in \mathcal{A}_0(x, y)} E_0^{x, y}\left[\int_0^{\infty} e^{-\beta t} U\left(c_t\right) d t\right]
\end{equation}
with the value function $\varphi^{*}(x,y)$ defined by
$$\varphi^{*}(x,y):= \sup _{(L, M, c) \in \mathcal{A}_0(x, y)} E_0^{x, y}\left[\int_0^{\infty} e^{-\beta t} U\left(c_t\right) dt\right].$$
Considering $w^*\left(x/y\right):=-\ln y+\beta \varphi^*(x, y)$, it is proved that $w^{*}(x)\in C^2((-1+\mu,\infty))/\{0\}$ satisfies the following variational problem in classical sense, see \citet{Shreve1994}: 
\begin{equation}\label{1d HJB}
\min \left\{-\mathcal{L}_0 w^*+\mathcal{A}^*_0 w^*, \frac{1}{x+1-\mu}-w^*_x, w^*_x-\frac{1}{x+1+\lambda}\right\}=0, \quad x>\mu -1 ,
\end{equation}
where
$$
\mathcal{A}^*_0 w^*=\beta\left(1- \ln{\beta} +w^*+\ln w^*_x\right) ,
$$
with free boundaries $x_s^* $ and $x_b^*$ such that 
\begin{equation}
	\begin{aligned}
& \mathbf{S R}^*:=\left\{x>\mu-1: w^*_x(x)=\frac{1}{x+1-\mu}\right\}=\left\{x>\mu-1: x\leq x^*_s \right\}, \\
& \mathbf{B R}^*:=\left\{x>\mu-1: w^*_x(x)=\frac{1}{x+1+\lambda}\right\}=\left\{x>\mu-1: x\geq x^*_b \right\}, \\
& \mathbf{N T}^*:=\left\{x>\mu-1: \frac{1}{x+1+\lambda}<w^*_x(x)<\frac{1}{x+1-\mu}\right\}=\left\{x>\mu-1: x_s^* < x < x^*_b \right\} .
\end{aligned}
\end{equation}

It is not hard to check that $v^*:=w^*_x$ is the solution to the following double obstacle problem:
\begin{equation}\label{double ob pro infi}
\begin{cases}-\mathcal{L}_1 v^*+\mathcal{A}^*_1 v^*=0 & \text { if } \frac{1}{x+1+\lambda}<v^*<\frac{1}{x+1-\mu}, \\ -\mathcal{L}_1 v^*+\mathcal{A}^*_1 v^* \leq 0 & \text { if } v^*=\frac{1}{x+1-\mu}, \\ -\mathcal{L}_1 v^*+\mathcal{A}^*_1v^* \geq 0 & \text { if } v^*=\frac{1}{x+1+\lambda},  \end{cases}
\end{equation}
where
\begin{equation*}
	\begin{aligned}
		&\mathcal{A}^*_{1} v^*(x,t)=	\beta\left(v^*+\frac{v^*_x}{v^*}\right).  
	\end{aligned}
\end{equation*}

With the above definition, we have the following result:
\begin{lemma}\label{lem xb posi}
If $\beta<1$, then $x_b(t)\geq x^*_b$, $\forall\ t\in [0, T]$. 
\end{lemma}
Therefore, when $\theta \geq \bar{\theta}$, we have 
$$x_s(t)\leq 0 \leq x_b(t), t \in [0, T].  $$ 

\begin{proof}[Proof of Lemma \ref{lem xb posi}]
We only need to show $v(x, t) \geq v^*(x)$.
Denote
$$
	\mathcal{N}=\left\{(x, t) \in \Omega_T \mid v^*(x) - v(x, t) > 0\right\}.
	$$
Then we have $ v^*(x) > \frac{1}{x+1+\lambda}$ and $ v(x, t) < \frac{1}{x+1-\mu}$ for $(x, t) \in\mathcal{N}$. Subsequently, 
\begin{align}
\begin{cases}
-v_{t}-\mathcal{L}_1 v+ \frac{1}{g(t)}\left( v+\frac{v_x}{v}\right)\geq 0, & \forall (x, t) \in\mathcal{N},\\
-\mathcal{L}_1 v^*+ \beta \left( v^* +\frac{v_x^*}{v^*}\right) \leq 0, & \forall (x, t) \in\mathcal{N}.
\end{cases}
\end{align}
In $\mathcal{N}$, $P(x, t):= v^*(x)- v(x, t)$ satisfies
\begin{align}
-P_t -\mathcal{L}_1 P +\beta \left( P+ \frac{P_x}{v^*}- \frac{v_x P}{v^*v} \right) + \left(\beta-\frac{1}{g(t)}\right)\left( v +\frac{v_x}{v}\right) \leq 0. 
\end{align}
Since $\beta \leq 1$, we have $g(t) \leq \frac{1}{\beta}$. Because $v +\frac{v_x}{v} \leq 0$ \citep[see][]{DAI20091445, Dai2009Consumption}, we derive
\begin{align}
-P_t -\mathcal{L}_1 P +\beta \left(P+ \frac{P_x}{v^*}- \frac{v_x P}{v^*v} \right) \leq 0. 
\end{align}
This implies $v^*(x)\leq v(x, t)$, and thus $\mathcal{N}=\emptyset$. Then $x_b(t)\geq x_b^*$ follows, otherwise
$$
v^*\left(x_b(t)\right)>\frac{1}{x_b(t)+1+\lambda}=v\left(x_b(t), t \right),
$$
which is a contradiction.
\end{proof}

We next prove the independence by contradiction.  
Assume that for some $\theta_1 \neq \theta_2$,
$$
\max_{x \in (-1, 0],\, t \in [0, T]} \{\hat{v}(x, t; \theta_1) - \hat{v}(x, t; \theta_2)\} = \delta >0.
$$
This maximum is achieved since the sell boundary exists, and as $x\to -1$, $\hat{v}(x, t; \theta_1)=\hat{v}(x, t; \theta_2)=\frac{1}{x+1}$.

(1) If the maximum $(x_0, t_0)$ lies on the boundary $x_0=0$, then
\begin{align*}
\hat{v}_{t}(x_0, t_0; \theta_1) \leq \hat{v}_{t}(x_0, t_0; \theta_2), \qquad
\hat{v}_{x-}(x_0, t_0; \theta_1) \geq \hat{v}_{x-}(x_0, t_0; \theta_2),
\end{align*}
and $\hat{v}(x_0, t_0; \theta_1) > \frac{1}{x+\theta_1}$ while $\hat{v}(x_0, t_0; \theta_2) < \frac{1}{x+1}$. Therefore,
\begin{align}
  -\hat{v}_{t}(x_0, t_0; \theta_1) - \mathcal{L}_1 \hat{v}(x_0, t_0; \theta_1) + \mathcal{A}_1 \hat{v}(x_0, t_0; \theta_1) \leq 0,\\
  -\hat{v}_{t}(x_0, t_0; \theta_2) - \mathcal{L}_1 \hat{v}(x_0, t_0; \theta_2) + \mathcal{A}_1 \hat{v}(x_0, t_0; \theta_2) \geq 0.
\end{align}
Thus $\hat{P}(x, t):=\hat{v}(x, t; \theta_1)-\hat{v}(x, t; \theta_2)$ satisfies at $(x_0, t_0)$
\begin{align*}
-\hat{P}_{t}-\mathcal{L}_1 \hat{P}+ \frac{1}{g(t)}\left(\hat{P} + \frac{\hat{v}_{x-}(\cdot; \theta_1)}{\hat{v}(\cdot;  \theta_1)}- \frac{\hat{v}_{x-}(\cdot; \theta_2)}{\hat{v}(\cdot; \theta_2)}\right) \leq 0 .
\end{align*}
Since $\hat{v}(x_0, t_0; \theta_1) > \hat{v}(x_0, t_0; \theta_2)\ge 0$ and $\hat{v}_{x-}(x_0, t_0; \theta_2)\le \hat{v}_{x-}(x_0, t_0; \theta_1)\le 0$, we get  
\[
\frac{\hat{v}_{x-}(x_0, t_0; \theta_1)}{\hat{v}(x_0, t_0; \theta_1)} - \frac{\hat{v}_{x-}(x_0, t_0;\theta_2)}{\hat{v}(x_0, t_0; \theta_2)} > 0,
\]
and hence
\[
-\hat{P}_{t}-\mathcal{L}_1 \hat{P}+ \frac{1}{g(t)}\hat{P} \leq 0.
\]
Since $\mathcal{L}_1 \hat{P}=-(\alpha-r-\sigma^2)\hat{P}$, we derive a contradiction from $\alpha-r-\sigma^2>0$, $g(t)>0$, and $\hat{P}_{t}\leq 0$.

(2) If the maximum is attained in the interior ($x_0<0$), then similarly
\[
-\hat{P}_{t}-\mathcal{L}_1 \hat{P}+ \frac{1}{g(t)}\hat{P} \leq 0
\]
in 
\[
\mathcal{N}:=\{(x,t)\in\hat{\Omega}_T: x<0,\ \hat{v}(x,t;\theta_1)>\hat{v}(x,t;\theta_2)\}.
\]
Since $g(t)>0$ and $\alpha-r\ge \sigma^2$, the maximum principle again gives a contradiction.

\section{\qsj{Additional numerical results}}\label{sec app posi merton}

\begin{figure}[htbp]
\begin{center}
\centering
\includegraphics[width=0.6\textwidth]{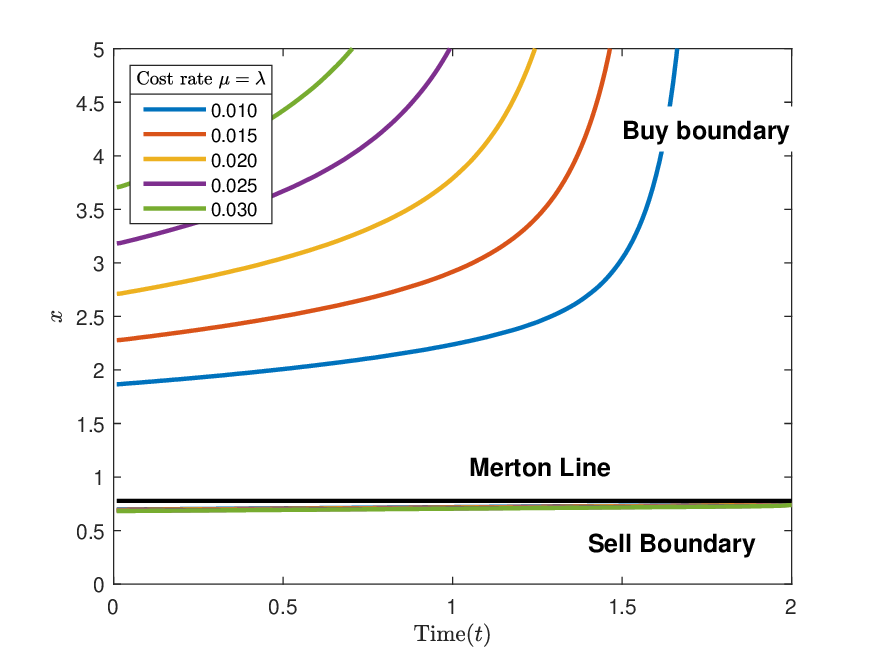}
\caption{\footnotesize Trading boundaries against time. Parameters: $T=2$, $\alpha = 0.1$, $r = 0.01$, and $\sigma = 0.4$. The corresponding Merton line is $x_M = 0.778$.}
\label{fig:tranding boundary with t posi}
\end{center}
\end{figure}

In this section, we plot figures for the case with a positive Merton line. We first present Figure \ref{fig:tranding boundary with t posi}. In comparison with Figure \ref{fig:tranding boundary with t}, the Merton line always lies within the no-trading region, which is consistent with Corollary \ref{corr mono cost}. 

\begin{figure}[htbp]
\begin{center}
\centering
\includegraphics[width=1\textwidth]{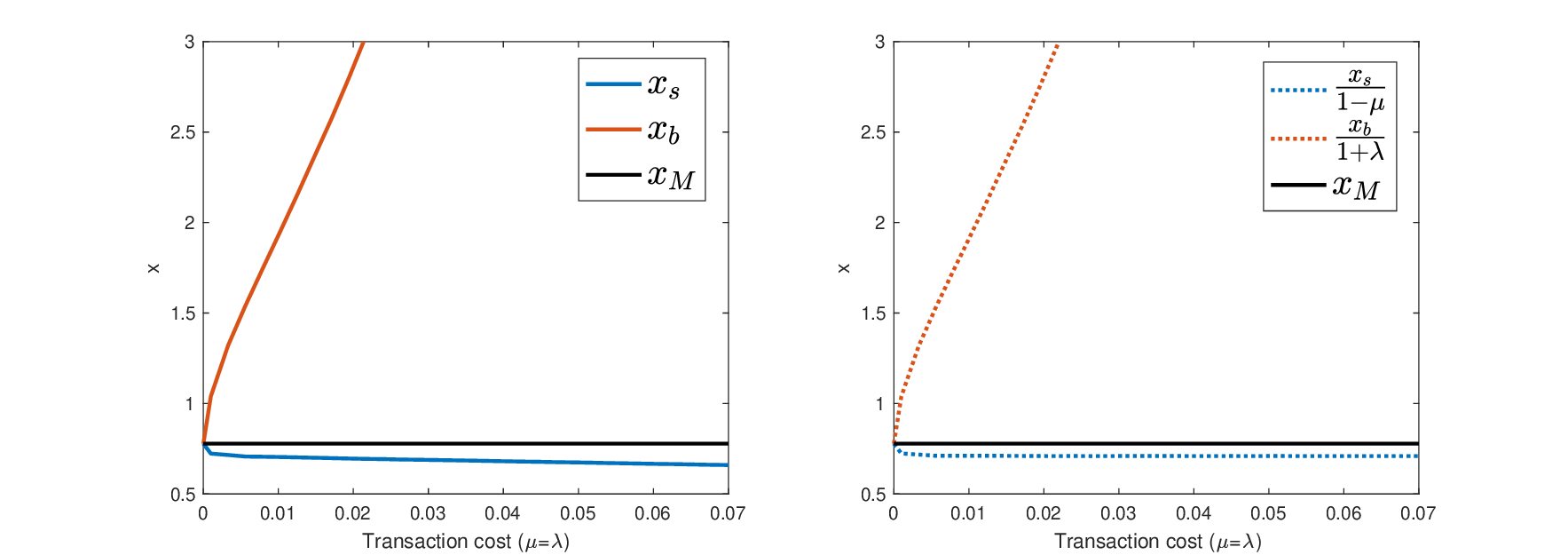}
\caption{\footnotesize Trading boundaries (left panel) and cost-adjusted trading boundaries (right panel) against transaction cost rates $\mu = \lambda$. Default parameters: $T=2$, $t=0.25$, $\alpha = 0.1$, $r = 0.01$, and $\sigma = 0.4$. The corresponding Merton line is $x_M = 0.778$.}
\label{fig:noFixedPositive}
\end{center}
\end{figure}

\begin{figure}[htbp]
\begin{center}
\centering
\includegraphics[width=1\textwidth]{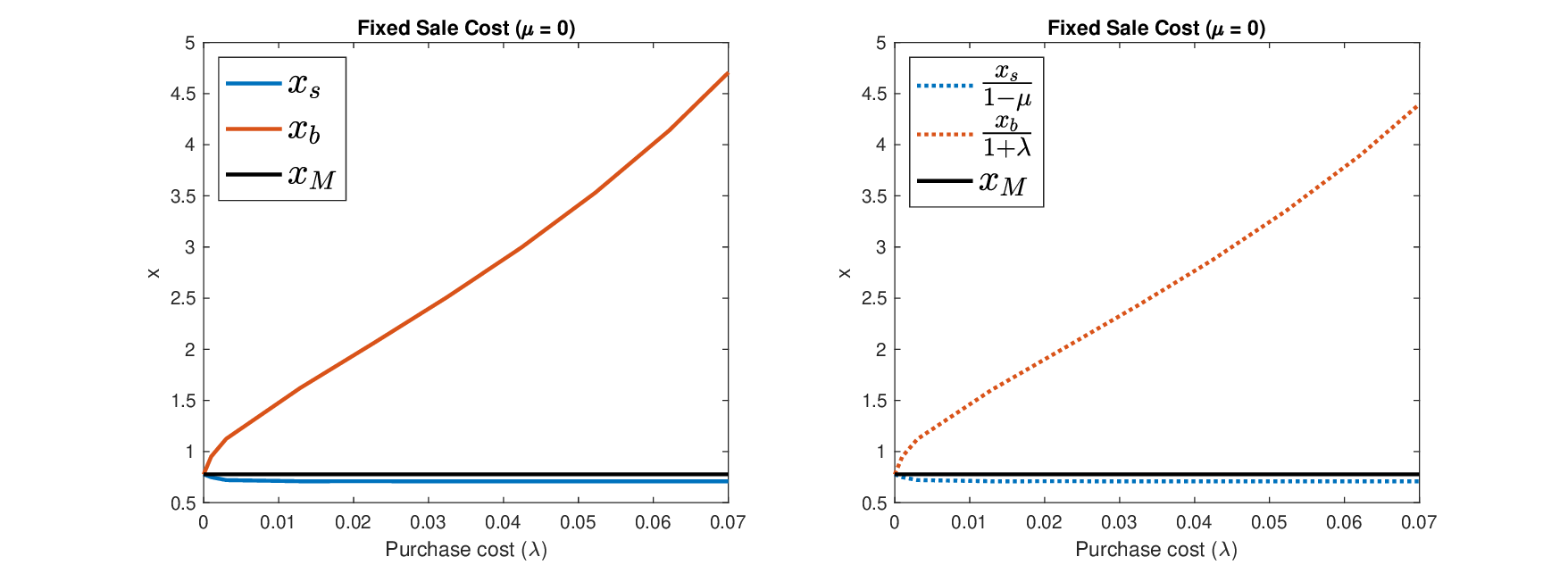}
\caption{\footnotesize Trading boundaries (left panel) and cost-adjusted trading boundaries (right panel) against purchase cost rate $\lambda$. Default parameters: $T=2$, $t=0.25$, $\alpha = 0.1$, $r = 0.01$, and $\sigma = 0.4$. The corresponding Merton line is $x_M = 0.778$.}
\label{fig:sellFixedPositive}
\end{center}
\end{figure}

\begin{figure}[htbp]
\begin{center}
\centering
\includegraphics[width=1\textwidth]{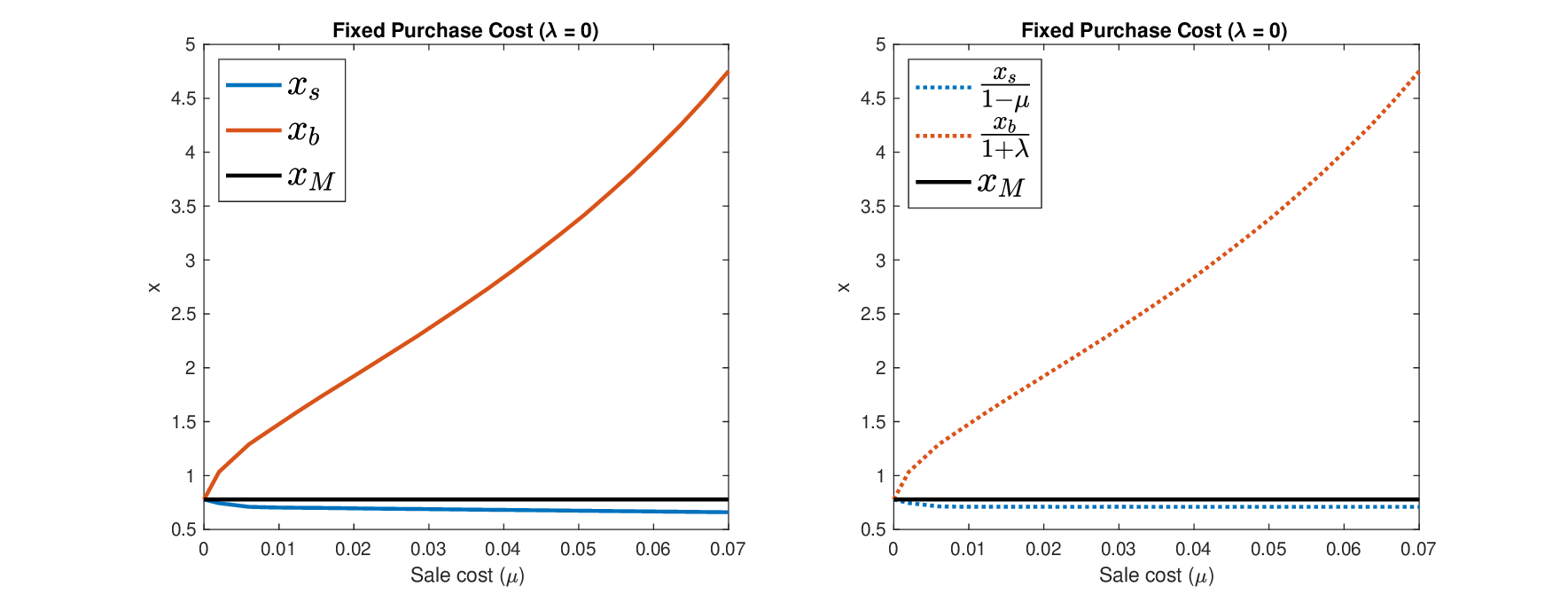}
\caption{\footnotesize Trading boundaries (left panel) and cost-adjusted trading boundaries (right panel) against sale cost rate $\mu$. Default parameters: $T=2$, $t=0.25$, $\alpha = 0.1$, $r = 0.01$, and $\sigma = 0.4$. The corresponding Merton line is $x_M = 0.778$.}
\label{fig:buyFixedPositive}
\end{center}
\end{figure}

In Figure \ref{fig:noFixedPositive}, we fix time $t$ and vary the transaction cost rates under the restriction $\mu = \lambda$. Compared to Figure \ref{fig:noFixedNegative}, the no-trading region in the left panel always contains the cost-adjusted no-trading region in the right panel. This is because in this case $x_b(t), x_s(t)>0$ by \citet{Dai2009Consumption}, and thus  
\[
x_s(t) < \frac{x_s(t)}{1-\mu} \leq x_M \leq \frac{x_b(t)}{1+\lambda} < x_b(t).
\]
In Figures \ref{fig:sellFixedPositive} and \ref{fig:buyFixedPositive}, we fix one transaction cost rate at zero and vary the other.

\clearpage
\section {Impact of other parameters }\label{Sec mono other}
In this section, we investigate the monotonicity of the trading boundary w.r.t. other parameters, using the approach in Section \ref{Sec mono cost}. We focus on the finite horizon case without consumption. 
\begin{theorem}\label{thm mono premium, sigma, gamma}
{For general CRRA utility $U(c) = \frac{c^\gamma}{\gamma}$ without consumption,} we have the following results. \\
(i). The sell boundary $x_{s}(t)$ and the buy boundary $x_{b}(t)$ are monotonically decreasing w.r.t. $\alpha-r$, the risk premium.\\
(ii). The sell boundary $x_{s}(t)$ and the buy boundary $x_{b}(t)$ are monotonically increasing w.r.t. $1-\gamma$.\\
(iii). When ${\frac{\alpha-r}{\sigma^2}}$ is fixed, the sell boundary $x_{s}(t)$ and the buy boundary $x_{b}(t)$ are monotonically decreasing w.r.t. $\sigma$.
\end{theorem}

Part (i) of Theorem \ref{thm mono premium, sigma, gamma} indicates that a higher risk premium induces {a higher proportion of wealth} in stock. Part (ii) {implies} that risk-averseness {reduces investment in stock}. Both {results} are consistent with {common sense}.
From Part (iii), we see  if one increases the volatility and risk premium simultaneously such that the Merton line, {the optimal proportion of wealth in stock without transaction costs}, is fixed, then the investor tends to invest  more in stock. {Because a higher volatility implies more adjustments and more transaction fees paid, thus the investor needs a higher portfolio return to compensate. }

\begin{proof}	
{We first prove (i) and (iii)}.
{For CRRA utility $U(c) = \frac{c^\gamma}{\gamma}$ without consumption, the corresponding double obstacle problem for $v(x, t): = w_x(x, t)$ is }
\begin{equation}
\left\{\begin{array}{l}
	-v_t-\mathcal{L}_1 v  =0\  \text{ if } \frac{1}{x+1+\lambda}<v<\frac{1}{x+1-\mu} , \\ 
	-v_t-\mathcal{L}_1 v \leq 0\   \text{ if } v=\frac{1}{x+1-\mu}, \\
	-v_t-\mathcal{L}_1 v \geq 0\  \text{ if } v=\frac{1}{x+1+\lambda} , \\
	v(x, T) =  \frac{1}{x+1-\mu},
\end{array}\right.\label{equ obstacleproblem ori}
\end{equation}
with 
$$
\begin{aligned}
\mathcal{L}_1 v(x,t)=&\frac{1}{2} \sigma^{2} x^{2} v_{x x}-\left(\alpha-r-(2-\gamma) \sigma^{2}\right) x v_{x}-\left(\alpha-r-(1-\gamma) \sigma^{2}\right) v\\
&+\gamma \sigma^{2}\left(x^{2} v v_x+x v^{2}\right).
\end{aligned}
$$
Before the proof of the results, we need the following lemmas.
	\begin{lemma}\label{lem esti vx}
		For the solution $v$ to the double obstacle problem \eqref{equ obstacleproblem ori}, we have $x v_x + v \geq 0$ in the whole region $\Omega_T$.	\end{lemma}

	\noindent{\bf Proof of Lemma \ref{lem esti vx}.}
		Define $\phi(x, t) = x v(x, t)$, then the lemma is equivalent to prove $\phi_x \geq 0$. {We have from \citet{Dai2009Consumption} that 
\begin{align}\label{equ vx esti}
-\frac{K}{(x+1-\mu)^2}\leq v_x \leq -v^2
\end{align}		
for some $K>0$. Then given $\frac{1}{x+1+\lambda}\leq v \leq \frac{1}{x+1-\mu}$, we have $\phi_x =x v_x +v >0$ in $ \{|x|< \delta \}$ for some small $\delta$. 
}		

Moreover, it is easy to verify that $\phi_x = x v_x + v \geq 0$ in the buying region and selling region. According to \eqref{equ obstacleproblem ori}, in $\mathbf{N T}\cap (\Omega_T\backslash \{|x|< \delta \})$ we have
		\begin{align}
			\phi_t+ \frac{1}{2}\sigma^2 x^2 \phi_{xx} - \bigg(\alpha -r -(1-\gamma)\sigma^2\bigg)x \phi_x +\gamma \sigma^2x \phi \phi_x = 0.
		\end{align}
	
	Given the uniform bound $|x|\geq \delta$, $\phi$ is actually a classical solution.  Then for function $\Phi := \phi_x$, we have 
		\begin{align}
			&\Phi_t+ \frac{1}{2}\sigma^2 x^2 \Phi_{xx} -(\alpha -r -(2-\gamma)\sigma^2)x \Phi_x \notag \\
			&\qquad -(\alpha -r -(1-\gamma)\sigma^2 ) \Phi + \gamma \sigma^2 \bigg( \phi\Phi + x \Phi^2 + x\phi\Phi_x  \bigg) = 0. \notag 
		\end{align}
		We have $\Phi\geq 0$ in the no-trading region by the comparison principle. \qed 
		\begin{lemma} \label{lem compare principle alpha}
	We let other parameters are fixed, and set $\frac{\alpha_1-r_1}{\sigma_1^2} \leq \frac{\alpha_2-r_2}{\sigma_2^2}$, $\sigma_{1} \leq \sigma_{2}$, then denote the corresponding solutions of the double obstacle problem \eqref{equ obstacleproblem ori} as $v_{1}$ and $v_{2}$, respectively. We have $v_{1} \geq v_{2}$ in $\Omega_T$.
	\end{lemma}
		\noindent{\bf Proof of Lemma \ref{lem compare principle alpha}.}
	Dividing \eqref{equ obstacleproblem ori} by $\sigma^{2}$, we have \begin{align}
&    -\frac{v_t}{\sigma^2} - \frac{1}{2} x^2 v_{xx} 
    + \left( \frac{\alpha - r}{\sigma^2} - (2-\gamma) \right) x v_x 
    + \left( \frac{\alpha - r}{\sigma^2} - (1-\gamma) \right) v \notag\\
 &\qquad  \qquad    - \gamma \left( x^2 v v_x + x v^2 \right) = 0, \quad \text{if } \frac{1}{x+1+\lambda} < v < \frac{1}{x+1-\mu}, 
\label{equ:boundary1}\\
&    -\frac{v_t}{\sigma^2} - \frac{1}{2} x^2 v_{xx} 
    + \left( \frac{\alpha - r}{\sigma^2} - (2-\gamma) \right) x v_x 
    + \left( \frac{\alpha - r}{\sigma^2} - (1-\gamma) \right) v  \notag\\
  &\qquad  \qquad    - \gamma \left( x^2 v v_x + x v^2 \right) \geq 0, \quad \text{if } v = \frac{1}{x+1+\lambda}, 
\label{equ:boundary2}\\
 &   -\frac{v_t}{\sigma^2} - \frac{1}{2} x^2 v_{xx} 
    + \left( \frac{\alpha - r}{\sigma^2} - (2-\gamma) \right) x v_x 
    + \left( \frac{\alpha - r}{\sigma^2} - (1-\gamma) \right) v \notag\\
   &\qquad \qquad    - \gamma \left( x^2 v v_x + x v^2 \right) \leq 0, \quad \text{if } v = \frac{1}{x+1-\mu}.
\label{equ:boundary3}
\end{align}

We prove the lemma by contradiction. Denote
		$$
		\mathcal{N}=\left\{(x, t) \in \Omega_T | v_{1}(x, t)<v_{2}(x, t)\right\}
		$$
		and assume $\mathcal{N}$ is a nonempty set. Then we must have 
		$$
		v_{1}<\frac{1}{x+1-\mu},\quad v_{2}>\frac{1}{x+1+\lambda} \text { in } \mathcal{N}.
		$$
		It follows that
		\begin{align*}
			&-\frac{v_{1t}}{\sigma_1^2}-\frac{1}{2} x^{2} v_{1xx}+\bigg(\frac{\alpha_1-r_1}{\sigma_1^2}-(2-\gamma)\bigg) x  v_{1x}\notag \\ 
			& \qquad \qquad +\bigg(\frac{\alpha_1-r_1}{\sigma_1^2}-(1-\gamma)\bigg) v_1 -\gamma \left(x^{2} v_1 v_{1x}+x v_1^{2}\right) \geq 0, \notag\\
			&-\frac{v_{2t}}{\sigma_2^2}-\frac{1}{2} x^{2} v_{2xx}+\bigg(\frac{\alpha_2-r_2}{\sigma_2^2}-(2-\gamma)\bigg) x  v_{2x}\notag \\ 
			& \qquad \qquad +\bigg(\frac{\alpha_2-r_2}{\sigma_2^2}-(1-\gamma)\bigg) v_2-\gamma \left(x^{2} v_2 v_{2x}+x v_2^{2}\right) \leq 0. \notag
		\end{align*}
		Let $u=v_{1}-v_{2}$, then we have in $\mathcal{N}$
		\begin{align}\label{equ comp1}
			& -\frac{u_{t}}{\sigma_2^2}-\frac{1}{2} x^{2} u_{xx}+\bigg(\frac{\alpha_2-r_2}{\sigma_2^2}-(2-\gamma)\bigg) x  u_{x}+\bigg(\frac{\alpha_2-r_2}{\sigma_2^2}-(1-\gamma)\bigg) u\notag \\
			& \qquad \qquad -\gamma \bigg(x^{2} (v_{1x} u + v_2 u_x) +x (v_1+v_2)u\bigg)\notag \\
			&\qquad \qquad + (\frac{1}{\sigma_2^2} - \frac{1}{\sigma_1^2})v_{1t} + (\frac{\alpha_1-r_1}{\sigma_1^2}-\frac{\alpha_2-r_2}{\sigma_2^2})(x v_{1x}+v_1)  \geq 0. 
		\end{align}
		\citet{DAI20091445}  show that\footnote{This inequality does not hold when incorporating consumption, see, {\citet{DaiZhong2008}}.} $v_{1t} \geq 0$  in $\Omega_T$.
Then combining \eqref{equ comp1} with Lemma \ref{lem esti vx}, we have 
		\begin{align}
			& -\frac{u_{t}}{\sigma_2^2}-\frac{1}{2} x^{2} u_{xx}+\bigg(\frac{\alpha_2-r_2}{\sigma_2^2}-(2-\gamma)\bigg) x  u_{x}+\bigg(\frac{\alpha_2-r_2}{\sigma_2^2}-(1-\gamma)\bigg) u \notag \\ 
			& \qquad \qquad -\gamma \bigg(x^{2} (v_{1x} u + v_2 u_x) +x (v_1+v_2)u\bigg) \geq 0. \notag
		\end{align}
		The comparison principle implies $u \geq 0$, contradiction. \qed

	Similar to the proof of Theorem \ref{thm: transaction con monotone}, we have (i) and (iii) with Lemma \ref{lem compare principle alpha}.

{To show (ii), we need} the following lemma. 
	\begin{lemma}\label{lem com v gamma}
		When other parameters are fixed, denote by $v_1$ and $v_2$ the corresponding solutions to problem \eqref{equ obstacleproblem ori} with relative risk aversion  $1-\gamma_1$ and $1-\gamma_2$, respectively.  If $1-\gamma_1 \geq 1-\gamma_2$, we have $v_1 \geq v_2$. 
	\end{lemma}
	\noindent{\bf Proof of Lemma \ref{lem com v gamma}. }
		Similar as the proof of Lemma \ref{prop compare transa cost}, if the set $\mathcal{N}: = \{(x,t) \in \Omega_T| v_1(x, t)< v_2(x, t)\}$ is nonempty, we have in $\mathcal{N}$
\begin{align}
&   -v_{1t} - \frac{1}{2} \sigma^2 x^2 v_{1xx} 
    + \left( \alpha - r - (2 - \gamma_1) \sigma^2 \right) x v_{1x} \notag\\
& \qquad \qquad     + \left( \alpha - r - (1 - \gamma_1) \sigma^2 \right) v_1 
    - \gamma_1 \sigma^2 \left( x^2 v_1 v_{1x} + x v_1^2 \right) \geq 0, 
\label{equ:v1_condition}\\
&    -v_{2t} - \frac{1}{2} \sigma^2 x^2 v_{2xx} 
    + \left( \alpha - r - (2 - \gamma_2) \sigma^2 \right) x v_{2x} \notag\\
& \qquad \qquad    + \left( \alpha - r - (1 - \gamma_2) \sigma^2 \right) v_2 
      - \gamma_2 \sigma^2 \left( x^2 v_2 v_{2x} + x v_2^2 \right) \leq 0.
\label{equ:v2_condition}
\end{align}
		Consequently, setting $u = v_1-v_2$, we have
\begin{align}
\begin{cases}
    -u_t - \frac{1}{2} \sigma^2 x^2 u_{xx} 
    + \left( \alpha - r - (2 - \gamma_2) \sigma^2 \right) x u_x 
    + \left( \alpha - r - (1 - \gamma_2) \sigma^2 \right) u \notag\\
 \qquad \qquad    - \gamma_2 \sigma^2 \bigg( x^2 (v_{1x} u + v_2 u_x) + x (v_1 + v_2) u \bigg) \notag\\
  \qquad \qquad    + (\gamma_1 - \gamma_2) \sigma^2 (1 - x v_1)(x v_{1x} + v_1) \geq 0, \quad \text{in } \mathcal{N}, \\
   u = 0, \quad \text{on } \partial_p \mathcal{N}.
\end{cases}
\end{align}
		Since 
\begin{align}
    &1 - x v_1 \geq 1 > 0, \quad \text{when } x \leq 0, 
\label{equ:condition_xv1_case1}\\
    & 1 - x v_1 \geq 1 - \frac{x}{x + 1 - \mu} = \frac{1 - \mu}{x + 1 - \mu} > 0, \quad \text{when } x \geq 0.
\label{equ:condition_xv1_case2}
\end{align}
		we have $u\geq 0$ in $\mathcal{N}$ from the comparison principle by noticing Lemma \ref{lem esti vx}. Contradiction.  \qed

	With Lemma \ref{lem com v gamma}, we have (ii) by comparing $v_1(x, t)$ and $v_2(x, t)$ as the proof of Theorem \ref{thm: transaction con monotone}.
\end{proof}

\subsection{\qsj{Numerical analysis}}
\begin{figure}[htbp]
\begin{center}
\centering
\includegraphics[width=0.6\textwidth]{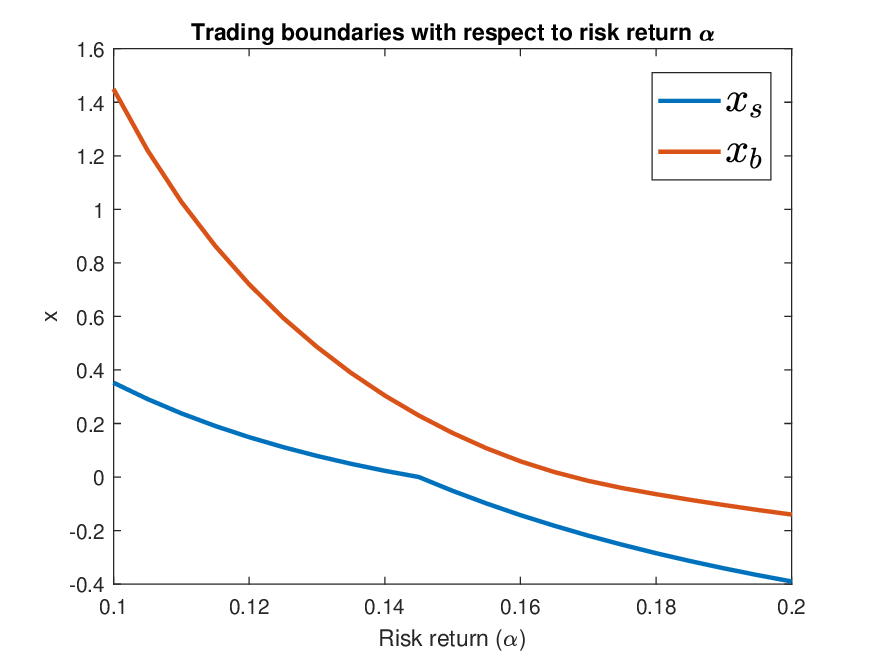}
\caption{\footnotesize Default parameters: $t=0.25$, $T=2$, $r = 0.02$, $\sigma = 0.5$, $\gamma=0.5$, $\beta=0.01$, and $\lambda = \mu = 0.02$.}
\label{fig:changeWithAlpha}
\end{center}
\end{figure}

\begin{figure}[htbp]
\begin{center}
\centering
\includegraphics[width=0.6\textwidth]{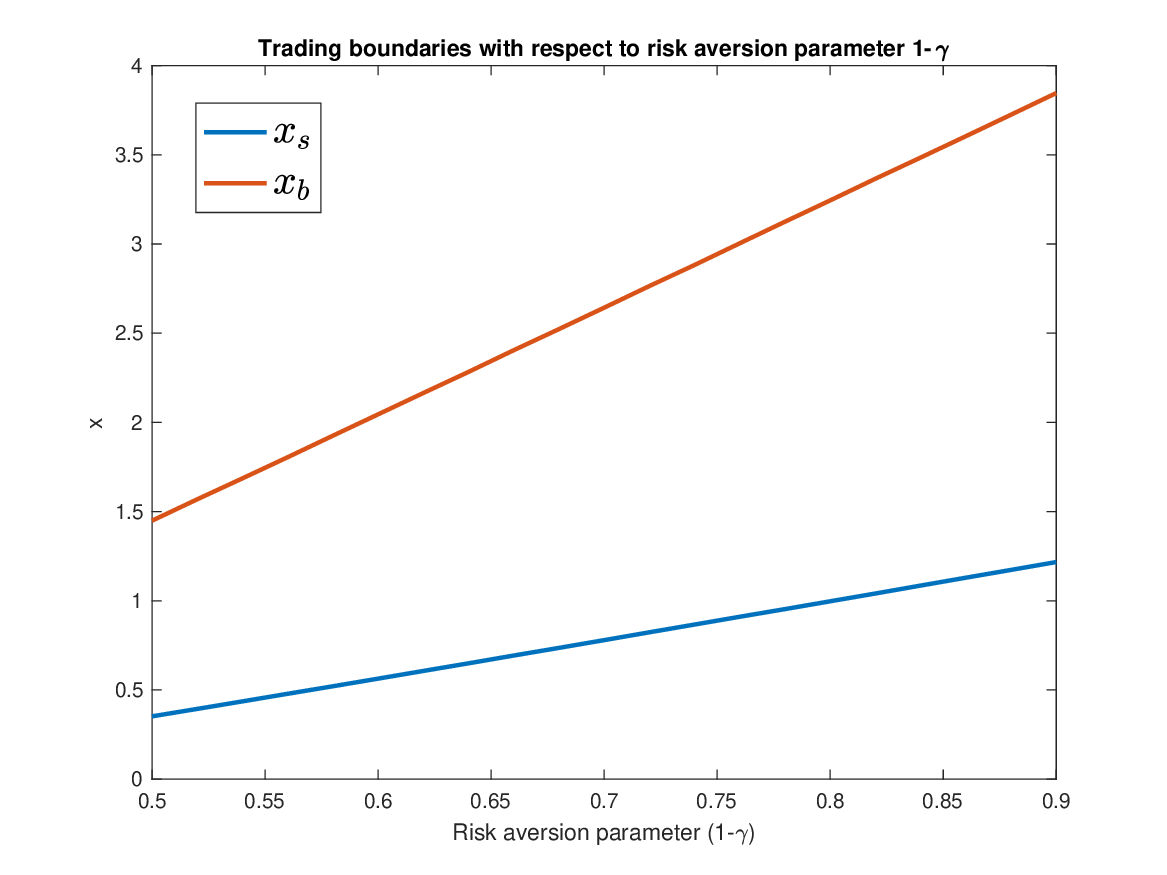}
\caption{\footnotesize  Default parameters: $t=0.25$, $T=2$, $\alpha=0.1$, $r = 0.02$, $\sigma = 0.5$, $\beta=0.01$, and $\lambda = \mu = 0.02$.}
\label{fig:changeWithGamma}
\end{center}
\end{figure}

\begin{figure}[htbp]
\begin{center}
\centering
\includegraphics[width=0.6\textwidth]{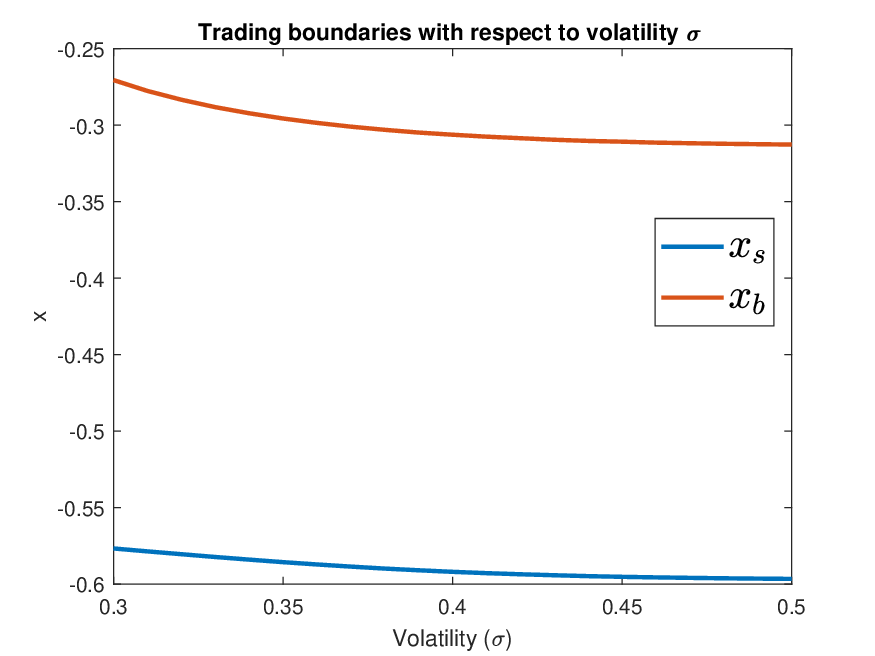}
\caption{\footnotesize Default parameters: $t=0.25$, $T=2$, $r = 0.02$,  $\frac{\alpha-r}{\sigma^2}=1$, $\beta=0.01$, and $\lambda = \mu = 0.02$.}
\label{fig:changeWithSigma}
\end{center}
\end{figure}

In Figures \ref{fig:changeWithAlpha}, \ref{fig:changeWithGamma}, and \ref{fig:changeWithSigma}, we numerically verify the results in Theorem \ref{thm mono premium, sigma, gamma}. A noticeable kink appears in the sell boundary when it crosses $x = 0$. The intuition is that the sell boundary remains positive (negative) for all $t$ when $\alpha - r - (1-\gamma)\sigma^2 < 0$ ($>0$); see \citet{DAI20091445} and \citet{Dai2009Consumption}. As $\alpha$ passes the critical value $r + (1-\gamma)\sigma^2$, the differential operator satisfied by the value function at $x = 0$ changes qualitatively, leading to the kink.

\end{appendices}

\bibliography{sn-bibliography}

\end{document}